\documentclass[11pt,a4paper]{article}

\usepackage[margin=1in]{geometry}
\usepackage[T1]{fontenc}
\usepackage{lmodern}
\usepackage[utf8]{inputenc}
\usepackage[protrusion=true,expansion=false]{microtype}
\usepackage{amsmath,amssymb}
\usepackage{booktabs}
\usepackage{array}
\usepackage{longtable}
\usepackage{tabularx}
\usepackage{graphicx}
\usepackage{xcolor}
\usepackage{enumitem}
\usepackage{listings}
\usepackage{caption}
\usepackage{placeins}
\usepackage[numbers,sort&compress]{natbib}
\usepackage{hyperref}

\hypersetup{
  pdftitle={HEPToolBench: Testing How Reliably Language Models Can Drive Particle Physics Software},
  pdfauthor={Aadarsh Singh and Sudhir K. Vempati},
  colorlinks=true,
  linkcolor=blue!60!black,
  citecolor=blue!60!black,
  urlcolor=blue!60!black
}

\newcommand{\heptoolbench}{\textsc{HEPToolBench}}
\newcommand{\proc}[1]{\texttt{#1}}

\title{\heptoolbench \hspace{1mm} 1.2: Testing How Reliably Language Models Can Drive Particle Physics Software}

\author{
Aadarsh Singh \quad Sudhir K.~Vempati\\[0.5em]
\small Centre for High Energy Physics,
Indian Institute of Science, Bengaluru, India\\[0.3em]
\small
\texttt{aadarshsingh@iisc.ac.in}
\quad
\texttt{vempati@iisc.ac.in}
}
\date{August 2026}

\begin{document}
\maketitle

\begin{abstract}
Scientists increasingly want to control research software through natural-language requests, but
fluent model output is useful only when it can be converted into a correct machine-readable
artifact. We introduce \heptoolbench{}, a benchmark of 28 collider-simulation tasks evaluated
using deterministic, task-specific scorers, together with a three-task structured-debugging
extension. We evaluate 42 model deployments, ranging from small locally served open source (open-weight)
models to hosted frontier systems. The central experiment compares direct generation of native
HEP-tool syntax with a schema-mediated interface in which the model returns a typed
representation and deterministic software performs serialization. Across five matched requests,
the mean score increases from 0.418 to 0.902 and task passes from 21/210 to 159/210, with 41
of 42 deployments improving; eleven deployments, seven of them locally served open-weight
models, move from no task pass under native syntax to five of five under the structured
interface. These values are measured relative to the corrected v1.2.1 benchmark contract, in
which the two native scorers that had enforced operational conventions absent from their own
prompts have been rescored prompt-faithfully across the whole cohort. One asymmetry remains in
the corrected contract: the native scorers score the recorded response verbatim and reject
responses wrapped in Markdown fences, whereas the structured scorers recover JSON from such
wrappers. A full-cohort rescoring under a single symmetric extraction rule has not been carried
out. An audit of the
archived responses of 17 locally served deployments shows that this remaining factor does not
account for the effect on that subset: task passes remain six to ten times more
frequent under the structured interface once both sides are scored under the same extraction
rule. A task pass does not by
itself guarantee runtime or scientific viability. Within this scope, the results show that moving
syntax generation into deterministic software can substantially improve the reliability of both
small local and frontier models. The prompts, scorers, model responses, and regeneration scripts
are released for independent evaluation and extension.

\end{abstract}

\section{Introduction}
\label{sec:introduction}
High-energy physics relies on specialist programs whose inputs are compact but unforgiving.
MadGraph5\_aMC@NLO~\cite{Alwall:2014hca}, Pythia~8~\cite{Sjostrand:2014zea,Bierlich:2022pix},
and Delphes~\cite{deFavereau:2013fsa}, for example, expect exact machine-readable commands,
cards, or configuration records. An error in one of these artifacts may stop the workflow
immediately or, more seriously, alter the simulated sample without producing an obvious failure.
An invalid antitop token prevents a MadGraph process from being generated, whereas entering the
total collision energy as the energy of each beam can produce a valid run at the wrong centre-of-mass
energy. Similarly, a cut applied at the wrong stage can change the selected event sample even when
the calculation completes successfully.
Natural-language agents offer a convenient interface to this software. Recent systems plan and
execute collider workflows, write analysis code, and attempt to reproduce published studies
\cite{Plehn:2026madagents,Diefenbacher:2025aod,Menzo:2025heptapod,Hill:2026grace,
Jiao:2026rooagent,Doglioni:2026agentrivet,PalaciosSchweitzer:2026colliderbench}. Their success,
however, depends on a lower-level question that can be hidden inside a larger orchestration system:
can the model produce the exact artifact required by the next program? A fluent explanation of the
physics does not guarantee a valid MadGraph command, run-card assignment, or structured record.
This question is particularly relevant to local deployment. Collaboration data and unpublished
results may need to remain within institutional infrastructure, large scans can turn small per-call
charges into a recurring cost, and compute nodes may have restricted external access.
Collaboration-specific assistants such as chATLAS and DUNE-GPT reflect the demand for systems that
can operate within experiment-specific infrastructure and data-governance constraints
\cite{DalSanto:2025chatlas,DUNEGPT:2026}. Open-weight models can also be pinned to a particular
version and served offline, but their practical value depends on whether the software interface
asks them to perform a task they can handle consistently.

\begin{figure}[!t]
    \centering
    \includegraphics[width=0.92\linewidth]{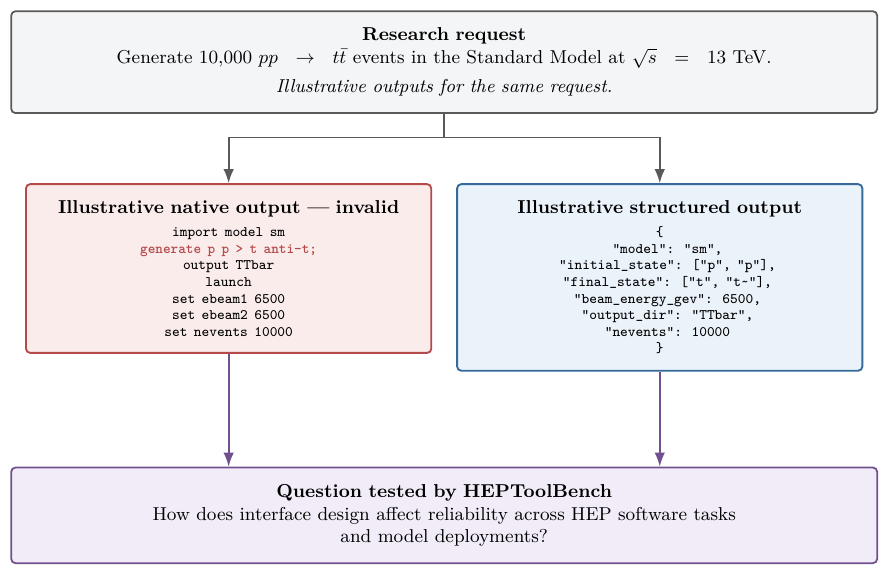}
    \caption{Illustrative comparison of native and structured interfaces for the same collider simulation request. The native response preserves the intended physics process but contains a malformed MadGraph line, whereas the structured response represents the request as named fields that deterministic software can convert into valid program input. This example shows one possible interface-level failure mode; HEPToolBench tests the broader effect of interface design on reliability across HEP software tasks and model deployments. The displayed outputs are illustrative rather than recorded benchmark responses.
}
    \label{fig:interface-example}
\end{figure}

\heptoolbench{} isolates this interface problem. Figure~\ref{fig:interface-example} shows it in a
simple form: a model may capture the intended physics but still produce an unusable command,
while a structured response leaves the exact tool syntax to deterministic software.
The version evaluated here contains 28
single-turn tasks covering process and card generation, workflow configuration, log and output
parsing, validation, diagnosis, scan planning and recovery, plot preparation, and reproducibility
auditing. A separate three-task extension evaluates the diagnosis and repair of supplied faulty
artifacts. Every task requires a concrete output and is graded by deterministic code; neither a
model judge nor a human preference score enters the evaluation.
The central experiment compares native tool syntax with a schema-described intermediate
representation. In the native-syntax condition, the model writes the final card or command
sequence. In the structured condition, it returns typed fields that deterministic code can validate
and render into the required tool syntax. Five matched task pairs keep the physics request and
the pass-critical physics requirements fixed while changing the interface presented to
the model. The comparison therefore tests a schema-mediated design---typed field decomposition,
structured extraction, and deterministic serialization---against direct native-syntax generation;
it does not separately identify the contribution of each element. The native side is scored under
the corrected v1.2.1 contract, in which the two native scorers that had required operational
conventions their own prompts never stated have been rescored prompt-faithfully. Because the
native and structured scorers still apply different extraction rules to the recorded response, we
audit how much of the measured difference that remaining asymmetry can explain
(Sec.~\ref{sec:scoring-policy}).

We evaluate 42 deployments representing 41 distinct model checkpoints, using both local and hosted
serving routes. The central result concerns the interface rather than the ranking of individual
models. On the five matched pairs, the structured representation improves the overall score for
41 of the 42 deployments. Several locally served open-weight models move from no task passes
under the native-syntax interface to four or five passes under the structured interface. Models with larger disclosed parameter counts generally perform better, but parameter count alone does not determine performance. These results show that much of the difficulty associated with native HEP syntax can
be moved into deterministic software, although validation of the resulting physics remains
necessary.

Section~\ref{sec:related} places \heptoolbench{} in the context of tool-using language models,
scientific-agent benchmarks, and recent HEP agents. Section~\ref{sec:benchmark-design} defines the
tasks, interface conditions, scoring policy, model cohort, and evaluation protocol.
Section~\ref{sec:results} presents the overall results, matched interface comparison, structured
debugging study, and run-to-run stability analysis. Section~\ref{sec:discussion-limitations}
discusses the implications and limitations of these results, and
Section~\ref{sec:conclusion} summarizes the conclusions. The benchmark, scorers, archived
responses, and regeneration scripts are released for independent checking and extension
\cite{HEPToolBenchRepo}.

\section{Related work}
\label{sec:related}
Tool-using language models are evaluated at several different levels. ReAct demonstrated how
reasoning and external actions can be interleaved, while Toolformer trained a model to decide when
and how to call external APIs~\cite{Yao:2022react,Schick:2023toolformer}. Subsequent benchmarks
test different parts of this process: the Berkeley Function Calling Leaderboard evaluates the
generation of function calls, SWE-bench asks models to resolve issues in software repositories,
and GAIA tests general assistants on questions requiring reasoning, information retrieval, and tool
use~\cite{BFCL:2024,Jimenez:2023swebench,Mialon:2023gaia}. \heptoolbench{} focuses on a narrower
unit of evaluation: whether one model response is already a usable scientific artifact before
execution feedback, retries, or repair by an agent scaffold.

Scientific coding and agent benchmarks have increasingly emphasized executable outputs.
SciCode evaluates code generation for research problems, ScienceAgentBench uses tasks drawn from
data-driven scientific workflows, and CORE-Bench tests whether agents can reproduce computational
results from published studies~\cite{Tian:2024scicode,Chen:2024scienceagentbench,
Siegel:2024corebench}. These benchmarks establish the broader context for evaluating scientific
work through concrete outputs rather than only natural-language answers.

Within HEP, agent systems now cover several stages of simulation and analysis. MadAgents operates
MadGraph and performs autonomous simulation campaigns, while HEPTAPOD connects language models to
schema-validated tools in multi-stage Monte Carlo workflows
\cite{Plehn:2026madagents,Menzo:2025heptapod}. ColliderAgent and SMEFT-Pheno-Agent address end-to-end collider-phenomenology workflows, and GRACE focuses on simulation-driven experiment design~\cite{Qiu:2026collideragent,Guo:2026smeftpheno,Hill:2026grace}. Beyond collider applications, DarkAgents constructs multi-agent pipelines for theoretical astroparticle physics; its first implementation studies cosmological first-order phase transitions and fits their gravitational-wave signal to NANOGrav data~\cite{Lucente:2026darkagents}. Analysis-oriented systems include Agents of Discovery for anomaly detection, RooAgent for ROOT-based analysis, Dr.Sai for BESIII workflows, and AgentRivet for constructing Rivet routines from publications~\cite{Diefenbacher:2025aod,Jiao:2026rooagent,He:2026drsai,Doglioni:2026agentrivet}. HepScript is
particularly close to the interface question studied here because it introduces a constrained
domain-specific language between human or model intent and experiment software
\cite{Jiao:2026hepscript}. Collider-Bench, by contrast, evaluates agents on the reconstruction of
published LHC analyses~\cite{PalaciosSchweitzer:2026colliderbench}. \heptoolbench{} complements
these systems by removing multi-step execution and repair from the measured unit, grading each
response deterministically, and evaluating both compact open-weight models and hosted models.

No decoding constraint is imposed in this experiment: the schema is described in the prompt, and
the benchmark parser then extracts the returned artifact for task-specific scoring.
Constrained-generation methods, which restrict the tokens available during inference so that the
output obeys a formal grammar~\cite{Scholak:2021picard,Poesia:2022synchromesh,
Willard:2023outlines,Dong:2024xgrammar}, would define a further interface condition that is not
evaluated here.

\section{Benchmark and evaluation}
\label{sec:benchmark-design}
\subsection{Tasks and interface conditions}
\label{sec:task-families}

A \heptoolbench{} task consists of a natural-language request, any input material needed to answer
it, a required output artifact, and a deterministic scorer. The model receives one user turn and
is instructed to return the artifact itself rather than an explanation of how it could be
constructed. Supplied inputs include short logs, cut-flow tables, SLHA excerpts, output manifests,
and scan records. The suite therefore tests both artifact generation and the interpretation of
existing workflow material, covering commands and cards, configuration, parsing, validation,
diagnosis, scan recovery, plot-ready data, and reproducibility records.
The version evaluated here contains 28 tasks in the main suite and a separate three-task
structured-debugging extension. The extension mirrors the three native-syntax card-repair tasks
in the main suite, but requires a JSON repair patch that identifies the error location and class
and supplies corrected content. We report the extension separately because it requires the model not only to correct a supplied
faulty artifact, but also to identify and classify the error in a structured repair patch. The
corresponding native-repair tasks require only the corrected artifact, so the two forms do not have
identical pass criteria and are not treated as controlled interface pairs. Table~\ref{tab:task-families} summarizes the task families; all task identifiers
and output types are listed in Appendix~\ref{app:tasks-scoring}.
\begin{table}[t]
\centering
\small
\begin{tabular}{p{0.32\textwidth} c p{0.44\textwidth}}
\toprule
Family & Tasks & Representative output \\
\midrule
Process generation (matched) 
& 6 (3 pairs) 
& Native MadGraph process artifacts and schema-described process objects \\

Native-syntax card repair 
& 3 
& Corrected MadGraph process cards \\

Run and workflow configuration (matched) 
& 4 (2 pairs) 
& Run-card settings and an MG5--Pythia~8--Delphes workflow in native and structured forms \\

Log and output parsing 
& 4 
& Run summaries, unit conversion, failure diagnosis, and output validation \\

Configuration and analysis validation 
& 4 
& Pythia, Delphes, LHE, and cut-flow diagnostics \\

Scan planning and recovery 
& 4 
& Scan grids, parameter-card patches, summaries, and rerun plans \\

Downstream decisions 
& 3 
& Benchmark selection, plot data, and a reproducibility audit \\

Structured debugging extension 
& 3 
& JSON repair patches with an error location, error class, and corrected content \\
\bottomrule
\end{tabular}
\caption{Task families in the version evaluated here. The first seven rows form the 28-task main
suite, while the final row is the separate three-task debugging extension. The five matched pairs
comprise three process-generation pairs, one run-card pair, and one workflow pair.}
\label{tab:task-families}
\end{table}

Eight main-suite tasks require native tool syntax, while twenty require a schema-described JSON
object. Comparing these two groups directly is useful as a description of the complete suite, but
it cannot identify an interface effect because they contain different tasks. The controlled
comparison instead uses five matched pairs: Drell--Yan, top-pair, and Higgs-plus-jet artifact
generation, a run-card request, and an MG5--Pythia~8--Delphes workflow. Within each pair, the
scientific request and the pass-critical physics requirements (process, beam energies, event
count, and requested stages) are held fixed, while the required interface changes from native
syntax to a schema-described representation. The structured condition combines explicit field
decomposition, recoverable structured-output extraction, and deterministic serialization. The
matched comparison measures this combined schema-mediated interface and does not isolate the
contribution of each component.

Two qualifications applied to the native side of the pairs as originally released, and one of
them has since been removed. The native scorers require operational conventions that are not
physics: an exact output directory name, a bare \texttt{launch} command, and fixed bookkeeping
lines such as \texttt{analysis=OFF} and a terminating \texttt{done}. For the Drell--Yan,
Higgs-plus-jet, and run-card pairs, every pass-critical requirement is stated in the native
prompt. For the top-pair process task (\texttt{mg\_basic\_002}) and the workflow task
(\texttt{mg\_workflow\_005}), however, the released scorers charged conventions that the
corresponding native prompts never stated. Both scorers have been corrected in v1.2.1 and the
whole cohort has been rescored under them: \texttt{mg\_basic\_002} now accepts any syntactically
valid \texttt{output <directory>} command instead of the unstated literal name \texttt{TTbar},
and \texttt{mg\_workflow\_005} no longer treats the unstated \texttt{analysis=OFF} and
\texttt{done} conventions as pass-critical or score-bearing. Every prompt-stated requirement
retains its original weight and enforcement in both tasks. All native counts reported in this
paper are the prompt-faithful v1.2.1 values; after the correction, one pair rather than two has
0/42 native passes (Table~\ref{tab:pairs}). The second qualification remains part of the
corrected contract: the extraction rules still differ between the two conditions, since a
native response containing a Markdown code fence or a visible reasoning block fails the native
contract, whereas the structured scorers strip such wrappers before parsing.
Section~\ref{sec:scoring-policy} quantifies the effect of this remaining asymmetry on archived
responses; it does not remove the interface effect.
The structured prompts specify field names, types, and constraints without supplying the target
values. The requested process, beam energy, event count, and other values must still be obtained
from the natural-language request or the supplied input material. The schema therefore constrains
the form of the response without disclosing its required content. Appendix~\ref{app:tasks-scoring}
gives a schematic example of this prompt contract.

\subsection{Scoring and clean-run policy}
\label{sec:scoring-policy}

Each scorer returns a continuous score in $[0,1]$ and a separate binary pass
label. The score assigns task-specific partial credit, whereas the pass label
requires the subset of fields defined as critical for that task. A process
artifact may therefore receive credit for the correct initial and final states
but fail because the beam energy is wrong. Pass is not obtained by applying a
universal threshold to the continuous score. We report both quantities because
the mean score measures partial correctness, while the pass count records how
often every pass-critical requirement is satisfied.

For structured tasks, deterministic artifact extraction precedes semantic
evaluation. Each task uses fixed extraction rules that are applied unchanged
across all deployments. These rules may recover a JSON object from a Markdown
code fence or surrounding explanation while recording the formatting deviation. A
small number of analysis-oriented tasks can also recover explicitly named fields
from otherwise legible JSON-like output. No model-specific parsing exceptions are
used. Once the required information is recovered, its values are checked and,
where applicable, the corresponding card or workflow is rendered deterministically
and validated. Native-syntax tasks are checked directly for the required commands
and settings. The scorers accept only explicitly encoded equivalences and do not
repair invalid MadGraph syntax. Further details are provided in Appendix~\ref{app:tasks-scoring}.

\paragraph{Convention and extraction audit.}
Two features of the released contract favoured the structured condition: the
native scorers enforced operational conventions concerning launch syntax, output
naming, and required bookkeeping commands, and they score the recorded response
verbatim, whereas the structured scorers recover a JSON object from the response
before checking it. We audited their combined effect on archived responses of the
five matched pairs without any new model calls. A common raw-response archive was
available for 17 locally served deployments but not for the hosted API
deployments, so the audit covers 85 matched units rather than the full 210; for
the structured side it also uses the 250 matched-task responses of the five-repeat
stability study. The audit was performed against the v1.2 scorers, and the counts
in this paragraph are therefore v1.2 quantities; the convention findings are what
motivated the v1.2.1 correction described in Sec.~\ref{sec:task-families}, and
that correction has since been applied to the full cohort, so the conventions are
no longer a scoring asymmetry to be discounted after the fact.
Three findings follow. First, applying the three
convention normalizations used by the companion system (renaming the output
directory, replacing \texttt{launch <name>} by a bare \texttt{launch}, and
inserting the fixed bookkeeping lines) to the native responses changes no task
pass on its own: the native count on the audited subset stays at 2/85, because
every remaining failure involves a physics field (beam energy, process, event
count, or seed) that is wrong or absent, rather than a convention. Second,
stripping Markdown fences and reasoning blocks from the native responses before
scoring, as the structured scorers already do, raises the native count to 11/85,
and after that normalization every native artifact with correct physics content
passes. Third, on the structured side, only 20 of the 72 passes on the audited
subset, and 101 of the 198 passes in the stability archive, would survive a
requirement of bare JSON with no fence or surrounding text; these are the
responses marked by the auxiliary strict-format flag (\texttt{strict\_passed})
that all five matched structured scorers record. The symmetric comparisons are
therefore 2/85 against 20/85 when both sides are scored verbatim and 11/85
against 72/85 when both sides use tolerant extraction. Under either rule the
structured interface passes six to ten times more often on this subset, so
neither the conventions nor the asymmetric extraction accounts for the effect.
The extraction asymmetry remains part of the corrected v1.2.1 contract, so it
still inflates the headline ratio, and the reported 21/210 native count should be
read as a lower bound on what a fence-tolerant native scorer would return. A
full-cohort rescoring under a single symmetric extraction rule has not been
carried out and remains the outstanding measurement; the convention correction,
by contrast, has been applied to all 42 deployments.

\paragraph{Native error taxonomy.}
The audit above shows how many native failures survive a symmetric scoring rule,
but not why they fail. A native response can go wrong because the physics is
missing or incorrect, because it breaks the tool's syntax and formatting
contract, or both, and the interface argument hinges on which of these
dominates. We classified every native-syntax output in the cohort to find out.
The unit is one completed response to one native task: 336 outputs (42
deployments $\times$ 8 native tasks), of which 30 pass under the corrected
v1.2.1 contract and 306 fail. Raw model text is available for all 336 outputs.
Every archived response
was checked against its stored SHA-256 before classification, and no score or
pass label was changed by the classification itself.

Each failed output was labelled on two independent axes. A
\emph{physics/task-content} error means a prompt-stated requirement, such as a
particle or antiparticle, the process, a beam energy, an event count, a cut, or a
requested workflow stage, is wrong or absent in the returned artifact. An
\emph{interface} error means an invalid MadGraph command or keyword, a malformed
process expression, Markdown or prose in a file-only response, a wrong
output-directory name, missing non-physics bookkeeping, or an extraction
failure. A wrong \emph{token} for the correct particle, such as \texttt{anti-t}
or \texttt{tbar} in place of the antitop \texttt{t\textasciitilde}, is counted as
an interface error rather than a physics error: the intended particle content is
right, and only the syntax is wrong. A wrong physical target, such as an
incorrect beam energy, remains a physics error. An output can carry both kinds of error, since the two axes are
independent. The classification records observable compliance of the returned
text with the prompt and the benchmark contract, not what a model understood
internally, and the detailed families overlap by construction and must not be
summed.

\begin{figure}[t]
  \centering
  \includegraphics[width=\textwidth]{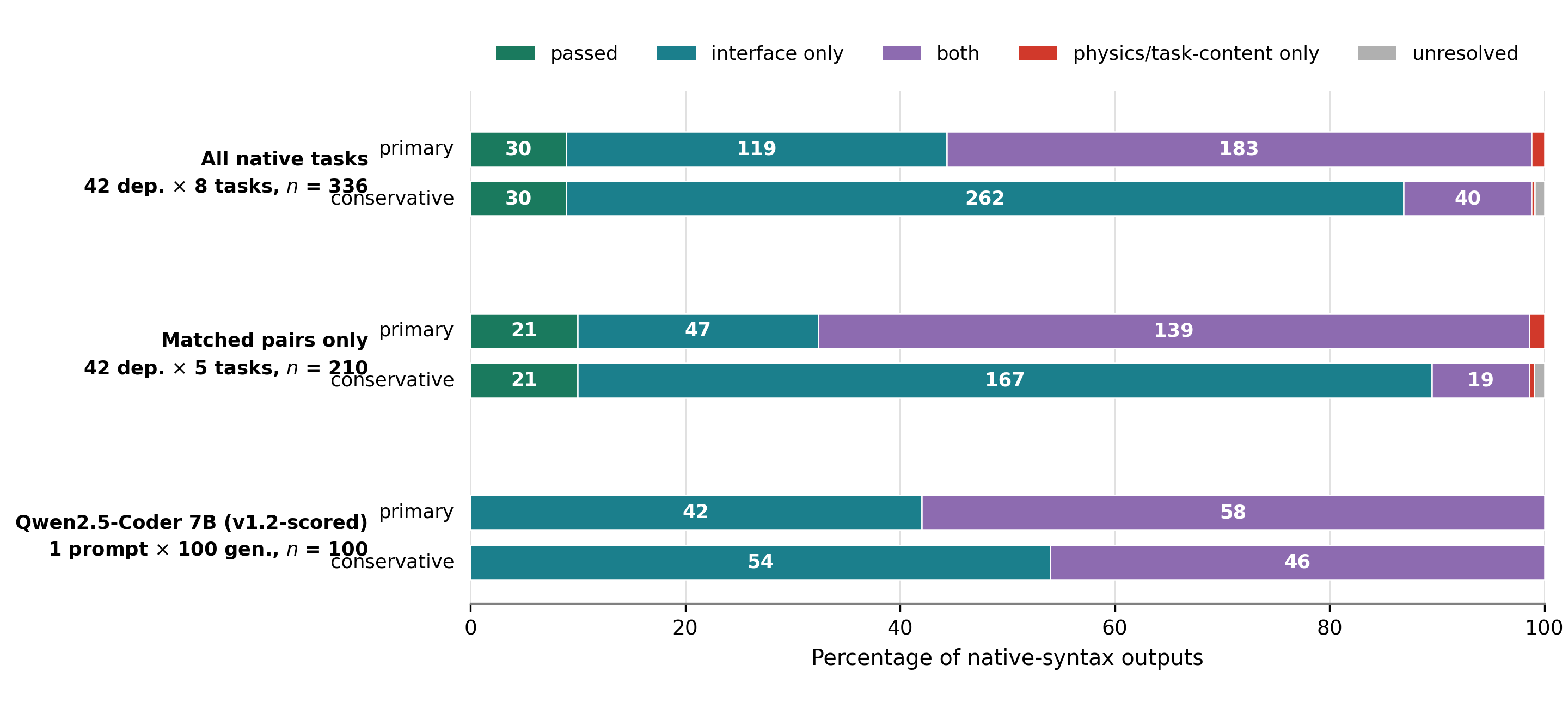}
  \caption{Composition of native-syntax errors. Each bar is one dataset under one
  mapping rule, divided into mutually exclusive categories; the number on each
  segment is an output count and the bar length is its percentage of that
  dataset. The \emph{primary} mapping charges an absent prompt-stated value to
  physics/task content, whereas the \emph{conservative} mapping counts only
  affirmatively wrong content. Top: all 336 native outputs (42 deployments
  $\times$ 8 native tasks). Middle: the 210 outputs of the five matched native
  tasks, whose 21 passes are the native count reported in
  Table~\ref{tab:pairs}. Bottom: 100 generations from a single
  deployment, \texttt{qwen2.5-coder:7b}, on one top-pair prompt, measuring
  within-model variability and not pooled with the cohort data; this probe alone
  is scored under the v1.2 contract, as noted in the text. The top and middle
  panels use the corrected v1.2.1 classification. An interface error is present
  in 302 of the 306 cohort failures under either mapping, while only four
  failures contain a physics/task-content error without an independent interface
  error.}
  \label{fig:native-error-taxonomy}
\end{figure}

One further choice affects the balance, so we report both settings. When a
required value is simply absent, the omission is ambiguous: the model may not
have supplied the physics, or it may have written an explanation instead of a
card. The \emph{primary} mapping charges every such omission to physics/task
content. This maximizes the physics share and is the setting least favourable to
an interface explanation. The \emph{conservative} mapping counts only content
that is affirmatively wrong and leaves pure omissions unassigned.
Figure~\ref{fig:native-error-taxonomy} shows the result. Of the 306 failed
outputs, an interface error is present in 302 (98.7\%) under either mapping,
while only four outputs in the entire cohort fail on physics content alone. What
moves between the mappings is the physics share: observable physics/task-content
errors appear in 187 failures (61.1\%) under the primary mapping and 41 (13.4\%)
under the conservative one. Restricting the classification to the five matched
native tasks reproduces the corrected native count of 21/210 and shows the same
pattern: 186 of the 189 failures (98.4\%) carry an interface error, and physics
involvement is 142 (75.1\%) or 20 (10.6\%) depending on the mapping. The largest
individual families are invalid MadGraph commands or keywords (184/336),
output-format violations (180/336), and missing prompt-stated content (168/336).
Table~\ref{tab:native-error-families} gives the full breakdown.

The taxonomy does not support a single explanation for all native failures. Under the primary mapping, which counts missing prompt requirements as content noncompliance, 187/306 failures (61.1\%) contain a physics/task-content error. Under the conservative mapping, only 41/306 (13.4\%) contain affirmatively wrong content. The robust result across both mappings is that 302/306 failures (98.7\%) contain an interface, syntax, or convention error, whereas only four failures contain a content error without an independent interface error.
These counts pool 42 deployments and eight requests, so they describe the
cohort, not any one model. To see the same split within a single model and a
single request, we ran \texttt{qwen2.5-coder:7b} on the top-pair native prompt
100 times with its serving configuration fixed, keeping every generation. The
prompt and all 100 responses were verified by SHA-256. This probe was scored
under the v1.2 contract and has not been rescored under the corrected
\texttt{mg\_basic\_002} scorer, so its counts are reported here as v1.2
quantities and are not pooled with the corrected cohort figures above. Under that
contract none of the 100 passes; every generation carries an interface error, and
physics/task-content errors appear
in 58 of them under the primary mapping and 46 under the conservative one. This
probe measures within-model variability for one checkpoint and one prompt. It is
not pooled with the 336-output dataset, and other checkpoints and other requests
should be expected to split differently between the two error types. This taxonomy is our own reading of the stored output against the prompt and the scorer's diagnostics, not an independent physics review, so the gap between the primary and conservative numbers reflects how much that judgment call matters. It also does not change any pass or fail count; that question is answered only by the audit above.

\paragraph{Clean-run policy.}
Infrastructure failures are treated as invalid measurements, not model failures.
Provider quota errors, server errors, timeouts, missing responses, and runner
exceptions are excluded and repeated, whereas a malformed artifact returned by a
completed model call remains a scored failure. A deployment enters the reported
cohort only after valid records are available for all 28 main-suite tasks and all
three extension tasks. The per-run metadata distinguish invalid calls, repeated
measurements, and completed responses that receive a score of zero. Further
details are provided in Appendix~\ref{app:tasks-scoring}.
\paragraph{Sarvam-105B repeated calls.}
Four Sarvam-105B calls returned a provider \texttt{HTTP 400} rejection with no
generated output. As provider-side rejections rather than model failures, they were
repeated under this policy using the same model identifier, prompts, decoding
settings, and API route as the canonical run. Because the provider is not
deterministic at \texttt{temperature}~$=0$, these are repeated measurements taken
after the canonical run rather than recovered text from the rejected requests;
per-record provenance is given in the released results. No HTTP-error placeholder
therefore remains in the cohort.

\begin{table}[t]
  \centering
  \caption{Native-syntax error families across the 336 native outputs (42
  deployments $\times$ 8 native tasks) under the corrected v1.2.1
  \texttt{B1\_W1\_S1R} cohort. Families are non-exclusive: one output may
  contain several errors, so the counts must not be summed. \emph{Local} and
  \emph{API} give the split over the 240 locally served and 96 hosted outputs. A
  wrong antitop token (\texttt{anti-t}, \texttt{tbar}) is counted under invalid
  process-expression syntax rather than as a wrong-particle error, following the
  taxonomy rule in the text.}
  \label{tab:native-error-families}
  \small
  \setlength{\tabcolsep}{5pt}
  \begin{tabular}{lccc}
    \hline
    Error family & Outputs & Local & API \\
    \hline
    \multicolumn{4}{l}{\emph{Physics/task-content}} \\
    \quad Wrong particle, antiparticle, process, or multiparticle definition & 31/336 & 27/240 & 4/96 \\
    \quad Wrong numerical collider or run setup & 12/336 & 8/240 & 4/96 \\
    \quad Wrong requested physics stage or workflow & 1/336 & 1/240 & 0/96 \\
    \quad Missing required prompt-stated physics/task content & 168/336 & 143/240 & 25/96 \\
    \multicolumn{4}{l}{\emph{Interface/syntax/convention}} \\
    \quad Invalid MadGraph command, keyword, or run-card syntax & 184/336 & 143/240 & 41/96 \\
    \quad Invalid process-expression syntax & 115/336 & 105/240 & 10/96 \\
    \quad Output-format violation, Markdown, prose, or chat text & 180/336 & 139/240 & 41/96 \\
    \quad Output-directory or benchmark naming convention & 75/336 & 63/240 & 12/96 \\
    \quad Non-physics bookkeeping or workflow-control requirement & 128/336 & 100/240 & 28/96 \\
    \quad Parser or extraction failure & 27/336 & 26/240 & 1/96 \\
    \multicolumn{4}{l}{\emph{Unresolved}} \\
    \quad Infrastructure failure with no model output & 0/336 & 0/240 & 0/96 \\
    \hline
  \end{tabular}
\end{table}
\label{sec:models-protocol}
\subsection{Model cohort and metadata}
\label{sec:cohort}

The final cohort contains 42 deployments representing 41 distinct model versions. A deployment
denotes a model version evaluated through a particular serving endpoint. Llama~3.3 70B is the only
model version represented twice: once through locally served Ollama inference and once through
GitHub Models. Thirty deployments use locally served models and twelve use hosted APIs. These labels
describe only the route through which inference was accessed. Four of the hosted deployments use
models whose weights are publicly downloadable, showing that serving route and weight availability
are separate properties.
We classify a deployment as open-weight only when weights for the corresponding model version are
publicly downloadable. API access or commercial private deployment alone does not meet this
definition. Under this rule, the cohort contains 34 open-weight and eight closed-weight deployments.
Several less obvious model labels affect this count. \texttt{mistral-large-2512} denotes Mistral
Large~3, an Apache-2.0 sparse mixture-of-experts model with 675 billion total and 41 billion active
parameters~\cite{Mistral:2025large3}. Similarly, \texttt{devstral-2512} denotes Devstral~2, a dense
123-billion-parameter model with publicly released weights~\cite{Mistral:2025devstral2}.
Sarvam-105B is an Apache-2.0 mixture-of-experts model with 105 billion total and 10.3 billion active
parameters~\cite{Sarvam:2026}. Codestral~25.08, by contrast, is listed by Mistral as a
``Premier'' model rather than an open-weight release and is therefore classified as closed-weight
under our definition~\cite{Mistral:2025codestral}.

Published total parameter counts are available for 34 deployments and range from 0.27 billion to
675 billion. For descriptive summaries, we divide the cohort into 25 deployments below 50 billion
parameters, nine at or above 50 billion, and eight whose total size is undisclosed. The 50-billion
boundary is a presentation convention rather than an inferred capability threshold. Quantitative
analyses use the disclosed total counts directly on a logarithmic axis.
Seven evaluated model versions use sparse mixture-of-experts architectures. For consistency with
the dense models, figures place them according to their total parameter counts, while published
active counts are also reported in the supplementary leaderboard. At a fixed numerical precision,
the total count is more closely related to the memory needed to store the model weights, whereas the
active count better reflects the expert computation used for each token.

\subsection{Evaluation protocol and computing environment}

For each deployment--task pair, the reported result is based on one completed model response.
The reported cohort therefore comprises 1176 main-suite observations (42 deployments $\times$ 28
tasks) and 1302 records in total once the three-task structured-debugging extension is included
(42 $\times$ 31), with exactly one record per deployment--task pair and no missing or duplicated
pairs. During inference, the model had no access to tools, execution feedback, or the scorer. A completed
response was retained and scored as returned, even when it was malformed or scientifically
incorrect; it was not repaired or repeated because of its content or score. Only calls invalidated
by an identifiable infrastructure failure could be repeated, following the clean-run policy in
Sec.~\ref{sec:scoring-policy}. The five-repeat study in Sec.~\ref{sec:stability} separately measures
the run-to-run variation associated with evaluating a single response.

Local inference for all 30 locally served deployments was performed on a Dell OptiPlex~7080
workstation running Manjaro Linux.\footnote{A similar exercise was also carried out on Apple
silicon hardware, using a Mac mini and a Mac Studio. Those runs are not part of the reported
cohort, which uses the single workstation described here.} The machine had an Intel
Core~i7-11700 CPU with eight physical
cores and 16 threads, 78\,GiB of RAM, Intel UHD~750 integrated graphics, and an NVIDIA
Quadro~P1000 with 4\,GiB of dedicated memory. Models were served with
Ollama~0.31.1~\cite{Ollama:2026}. Models that could not fit entirely in GPU memory therefore relied
substantially on the CPU and system memory. The benchmark runner used Python~3.12.3. Local
responses were collected through Ollama's non-streaming HTTP interface, saved before scoring, and
evaluated offline. The exact model labels used in the evaluation are listed in the supplementary
leaderboard.

Decoding configuration was held fixed throughout the evaluation. Hosted API deployments were
queried at temperature~0. Locally served deployments were run through Ollama in each model's default
serving configuration, except the two Qwen3.5 deployments (27B and 35B), which were run with
reasoning (``thinking'') disabled and should be read as distinct serving configurations rather than
equivalent inference modes; the reasoning mode and context window recorded for each deployment are
included in the released results. Locally served models were pulled and run using their standard
Ollama tags, that is, their default quantization. For the deployments whose model metadata was
captured, this quantization was \texttt{Q4\_K\_M} (four-bit $K$-quant) in every case except gpt-oss,
released and served in \texttt{MXFP4}; the parameter size and captured quantization per deployment
are included in the released results. Each response was bounded by its deployment's context
window, which ranged from 4096 to 16\,384 tokens across the cohort; a response that reaches this
limit terminates with a length stop reason that is recorded, and run-to-run variation under these
settings is characterized in Sec.~\ref{sec:stability}.
Only response content exposed by the serving endpoint was stored and scored; hidden provider
reasoning was neither requested nor available to the scorer. Visible reasoning or explanatory
text remained part of the recorded response and was handled by the same fixed extraction rules.
In practice this means that a native-syntax response containing a reasoning block or a Markdown
fence fails the verbatim native contract, whereas a structured response is parsed after such wrappers
are removed. Among the locally served deployments, separate reasoning content was recorded by the
serving endpoint only for gpt-oss~120B and Qwen3-Next 80B; the audit in
Sec.~\ref{sec:scoring-policy} quantifies how many native failures this rule produces.

\section{Results}
\label{sec:results}
\subsection{Overall capability and model scale}
\label{sec:overview}

\begin{figure*}[!t]
\centering
\includegraphics[height=0.82\textheight,keepaspectratio]{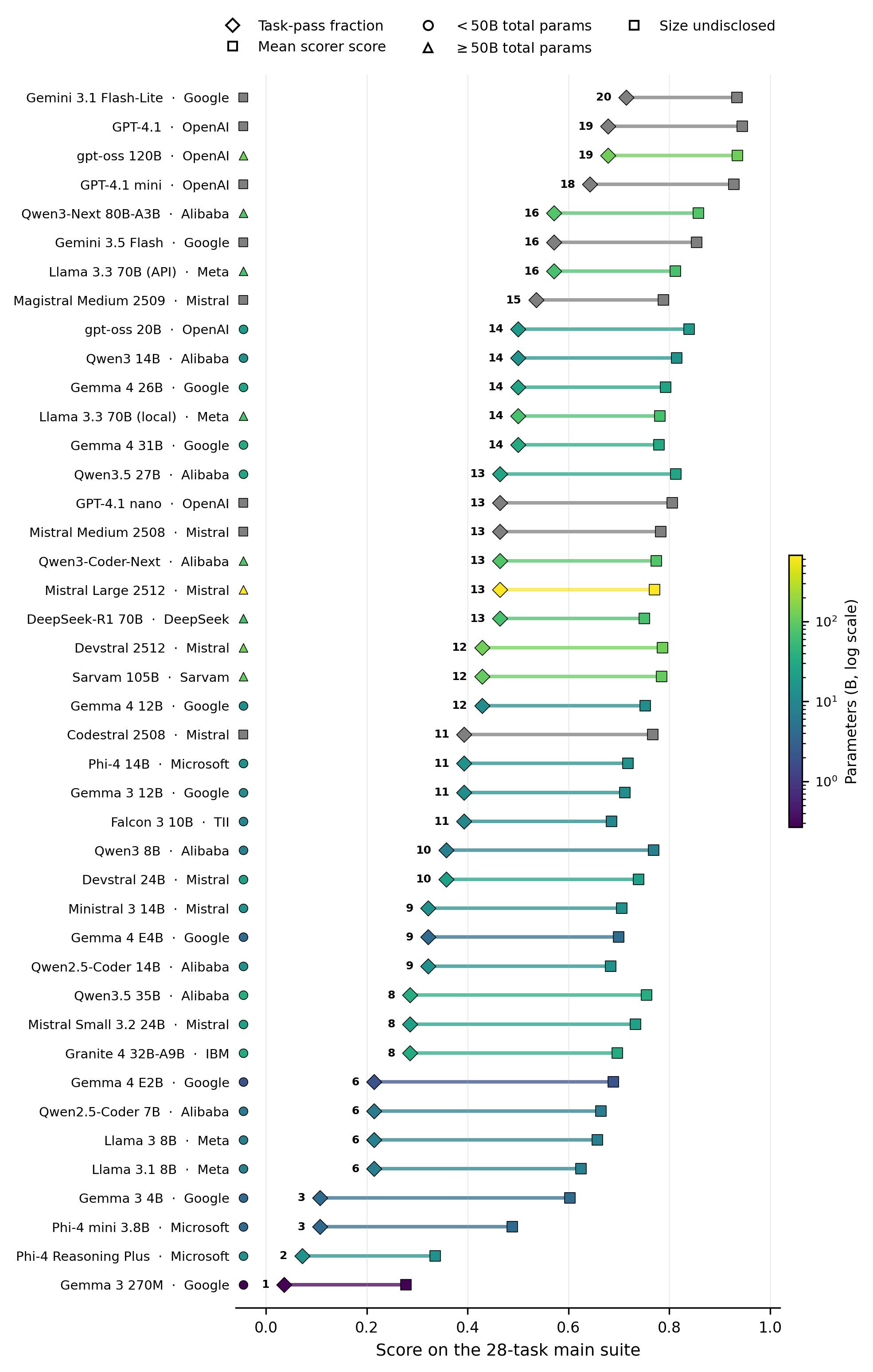}
\caption{Performance of all 42 deployments on the 28-task main suite, ordered by task-pass
fraction. Diamonds show the pass fraction, squares show the mean scorer score, and the number
beside each diamond gives the task-pass count. Colour encodes the disclosed total parameter
count on a logarithmic scale, while grey denotes an undisclosed size. The symbols beside the
model names distinguish deployments below 50 billion parameters, those at or above 50 billion,
and those with undisclosed size.}
\label{fig:overview}
\end{figure*}

Figure~\ref{fig:overview} compares all 42 deployments on the 28-task main suite. Task passes
range from 1/28 to 20/28, while mean scores range from 0.277 to 0.945. The suite therefore
separates the evaluated deployments without producing either a ceiling or a common floor.
Gemini~3.1 Flash-Lite passes 20 tasks, followed by GPT-4.1 and the open-weight, locally
served gpt-oss~120B with 19 each. Their mean scores are 0.934, 0.945, and 0.936, respectively.
Because these deployments differ by at most one task pass, and the highest pass count and the
highest mean score are not achieved by the same deployment, they are better regarded as the
leading group than as a precisely resolved ranking. The complete deployment-level results are
given in Appendix~\ref{app:model-results}.

\begin{figure}[t]
\centering
\includegraphics[width=0.98\textwidth]{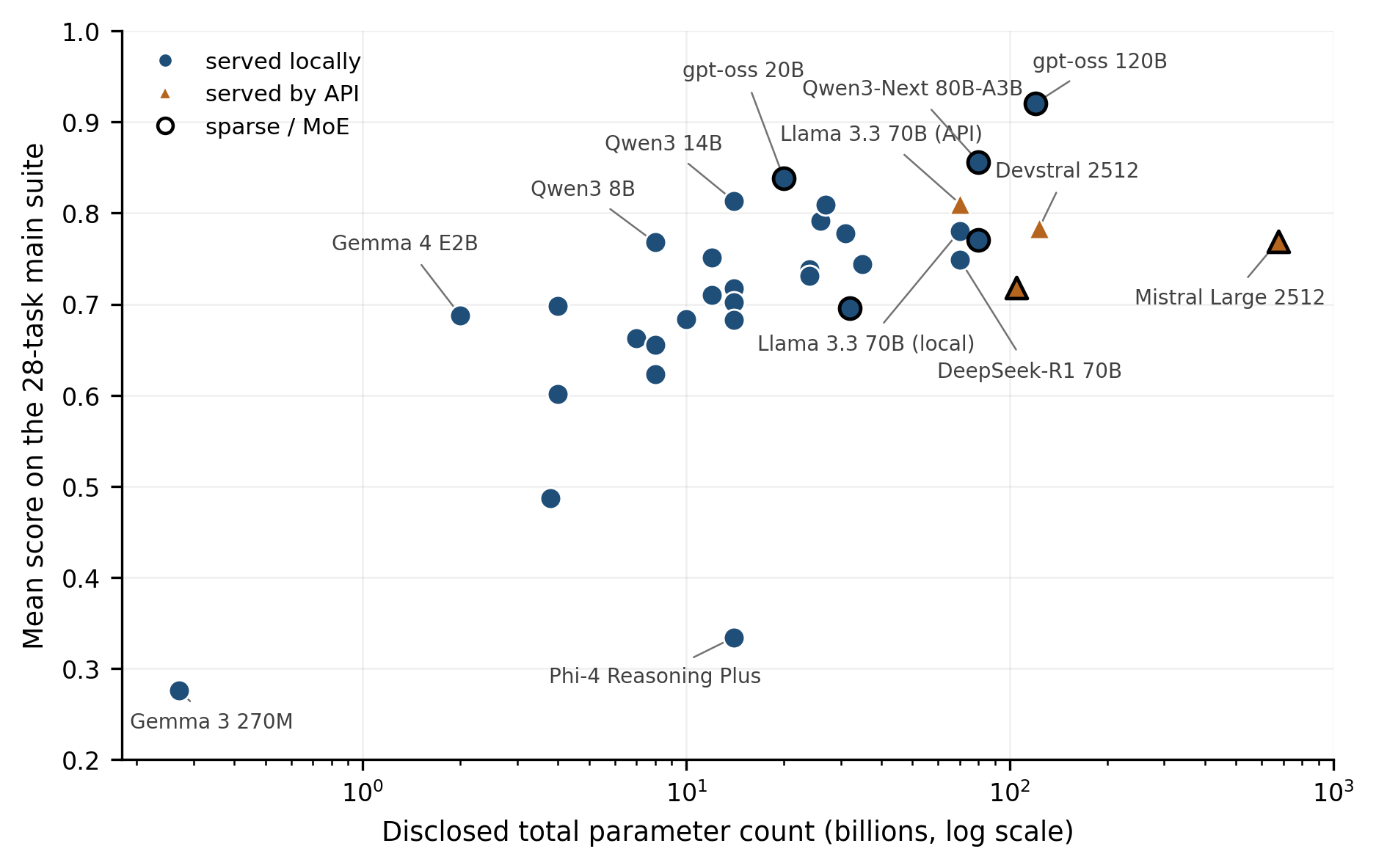}
\caption{Mean main-suite score versus disclosed total parameter count for 34 deployments.
Circles denote locally served deployments and triangles denote API-served deployments. A black
outline marks sparse mixture-of-experts models, which are positioned according to total rather
than active parameter count. The eight deployments with undisclosed size are not shown.}
\label{fig:scale}
\end{figure}

The two metrics provide complementary views of performance. Because each task has its own
pass-critical requirements, a task pass is not obtained by applying a universal threshold to the
continuous score. Across the main-suite responses, the highest failing score is 0.96 and the
lowest passing score is 0.79. Moreover, on 14 of the 28 tasks, no observed response with a score
below 0.95 passes. Mean score therefore summarizes partial correctness, whereas task-pass count
records how often all critical requirements are satisfied in the returned artifact.
Model size is associated with performance within this cohort, but does not determine it. Among the
34 deployments with a disclosed total parameter count, larger models generally achieve higher mean
scores. The descriptive Spearman correlation is $\rho=0.737$ ($p=6.8\times10^{-7}$, $n=34$), although the scatter in
Fig.~\ref{fig:scale} shows that parameter count alone does not determine performance. Substantial variation remains among models of the same reported size:
Qwen3~14B scores 0.815, whereas Phi-4 Reasoning Plus, also reported as a 14-billion-parameter
model, scores 0.336. Parameter count is therefore informative at the cohort level but is not
sufficient for selecting an individual model.

Serving route cannot be isolated from this comparison because model identity, weight availability,
size, provider, runtime, and quantization are not controlled independently. Llama~3.3 70B is the only model version evaluated through both serving routes. The locally
served deployment scores 0.782 and passes 14 tasks, while the GitHub Models deployment scores
0.812 and passes 16. Across five repeat runs of the local deployment, the pass count ranges from
13 to 16 and the mean score from 0.770 to 0.796. The API pass count therefore lies within the
observed local run-to-run variation, although its mean score is slightly higher than all five local
repeat scores. 

\subsection{The interface effect}
\label{sec:interface}

The strongest cohort-wide difference is associated with the required output interface. Across all
42 deployments, the eight native-syntax tasks yield 30 task passes in 336 model--task cases,
with a mean score of 0.443. The twenty schema-described tasks yield 441 passes in 840 cases, with
a mean score of 0.859. Thirty-one deployments pass none of the eight native-syntax tasks. Because
these two task sets contain different requests, however, this descriptive comparison does not by
itself establish an interface effect.

\begin{figure*}[!t]
\centering
\includegraphics[width=0.98\textwidth]{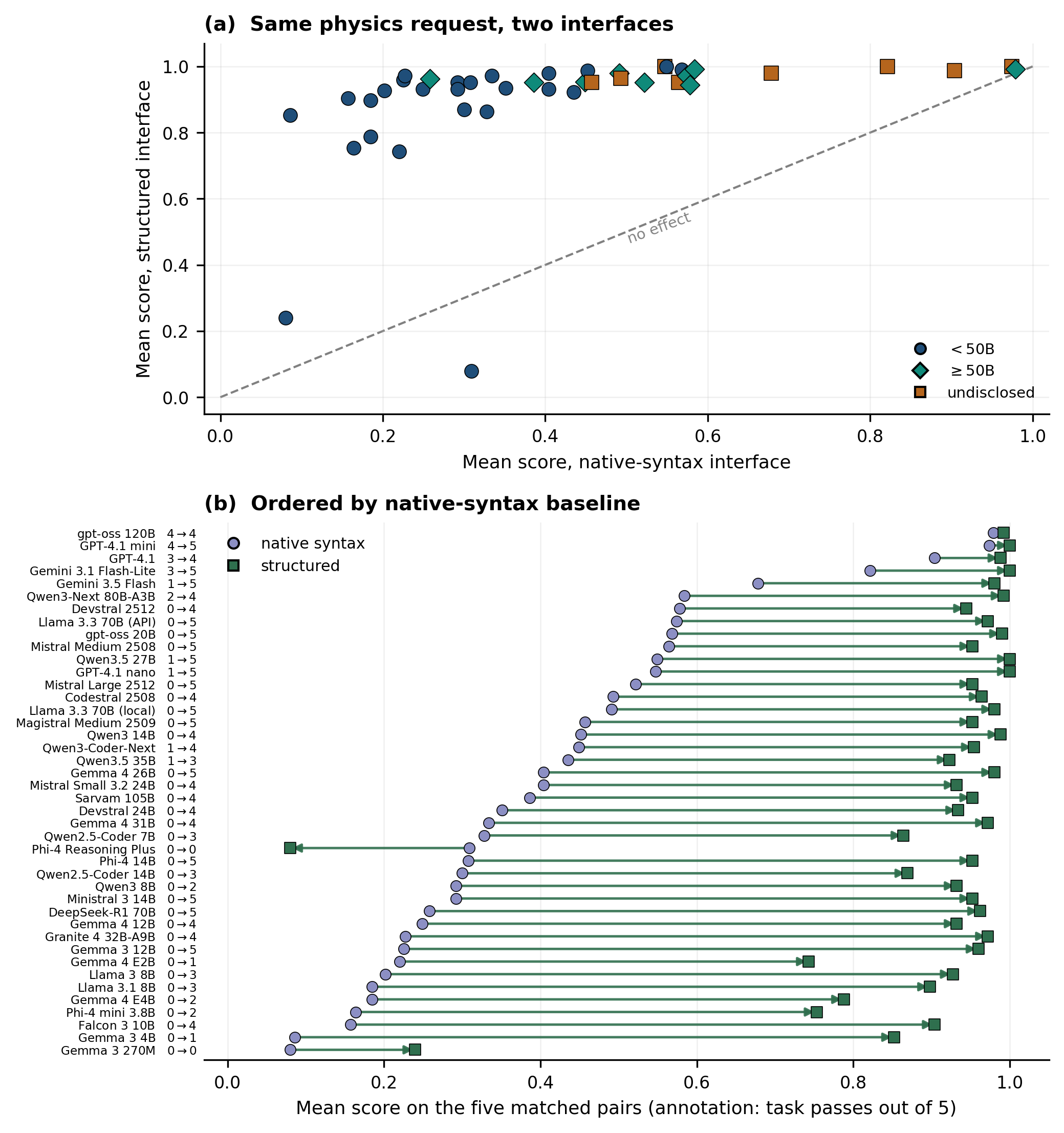}
\caption{Performance on the five matched native-syntax and structured task pairs.
\textbf{(a)} Each point represents one deployment; points above the diagonal perform better under
the structured interface. Circles, diamonds, and squares denote deployments below 50 billion
parameters, those at or above 50 billion, and those with undisclosed size, respectively.
\textbf{(b)} The deployments ordered by their native-syntax mean score; labels give the change in
task passes out of five.}
\label{fig:interface}
\end{figure*}

The controlled evidence comes from the five matched pairs, which keep the physics request and
the physics pass-critical checks fixed while changing the required artifact from native syntax
to schema-described JSON. The mean score rises from 0.418 to 0.902, and task passes increase from
21/210 to 159/210 (Table~\ref{tab:pairs} and Fig.~\ref{fig:interface}). The matched mean improves for 41 of the 42 deployments. A one-sided Wilcoxon signed-rank test
provides descriptive support for a cohort-wide shift in favour of the structured interface
($W=897$ out of a maximum of $903$, $p=3.2\times10^{-12}$). The test is applied to the 42
deployment-level pairs of matched means, one pair per deployment; the reported statistic is the
sum of the positive-signed ranks, $W^{+}$, and the $p$-value is the one-sided exact
permutation value from \texttt{scipy.stats.wilcoxon} with \texttt{alternative="greater"} and
\texttt{method="exact"}. No zero differences occur, so no zero-handling or continuity correction
enters. All $p$-values reported here for signed-rank tests are exact under this convention.
The improvement is therefore
widely shared across the evaluated cohort rather than being produced by only a few deployments.
Because the deployments include related model families and five fixed requests, this should not
be interpreted as population-level inference. As a check against the clustering of related
deployments, we also aggregated the matched means to one value per model family
\footnote{The twelve developer-lineage groups are OpenAI GPT-4.1, gpt-oss, Gemini, Qwen, Llama,
Mistral (including Devstral, Codestral, Magistral, and Ministral), Gemma, Phi, DeepSeek, Granite,
Sarvam, and Falcon.}
(twelve families defined by developer lineage): every family improves, the smallest family-level
gain is $+0.19$ (the GPT-4.1 family, which already has the highest native mean), and the
family-level Wilcoxon statistic attains its maximum ($W=78$, $p=2.4\times10^{-4}$, exact,
one-sided, under the same convention). All five
task pairs improve in both mean score and task-pass count, showing that the aggregate result is
not driven by a single request. After the v1.2.1 rescoring, one pair rather than two has no
native pass: the top-pair task moves from 0/42 to 4/42 once its scorer stops requiring an output
directory name that its prompt never states, leaving the MG5--Pythia~8--Delphes workflow as the
only matched request with no native pass in the cohort (Sec.~\ref{sec:task-families}).

\begin{table}[t]
\centering\small
\caption{The five matched interface pairs, aggregated over all 42 deployments, under the
corrected v1.2.1 \texttt{B1\_W1\_S1R} contract. The physics request
and the physics pass-critical checks are identical within each pair; the model supplies either
native syntax or a schema-described representation that deterministic software serializes. The
native scorers score the recorded response verbatim and reject fenced responses
(Sec.~\ref{sec:scoring-policy}); the operational conventions that the top-pair and workflow
prompts never stated are no longer charged. The single pair with 0/42 native passes is the
workflow pair, whose remaining failures are not convention failures.}
\label{tab:pairs}
\begin{tabular}{lrrrcc}
\toprule
& \multicolumn{3}{c}{Mean score} & \multicolumn{2}{c}{Task passes} \\
\cmidrule(lr){2-4}\cmidrule(lr){5-6}
Task pair & Native syntax & Structured & $\Delta$ & Native syntax & Structured \\
\midrule
Drell--Yan process & 0.505 & 0.901 & $+0.396$ & 6/42 & 35/42 \\
Top-pair process & 0.365 & 0.901 & $+0.536$ & 4/42 & 33/42 \\
Higgs+jet process & 0.471 & 0.930 & $+0.459$ & 6/42 & 35/42 \\
Run-card cuts & 0.345 & 0.930 & $+0.586$ & 5/42 & 39/42 \\
MG5+P8+Delphes workflow & 0.405 & 0.845 & $+0.440$ & 0/42 & 17/42 \\
\midrule
\textbf{All five pairs} & 0.418 & 0.902 & $\mathbf{+0.484}$ & 21/210 & \textbf{159/210} \\
\bottomrule
\end{tabular}
\end{table}

The improvement is not confined to the largest models. Among the 25 deployments below 50 billion
parameters, task passes increase from 2/125 to 81/125; the complete size-class breakdown is given
in Table~\ref{tab:sizeclass}. Eleven deployments move from passing none of the five native-syntax
tasks to passing all five structured counterparts. These include Gemma~3 12B, Ministral~3 14B,
Phi-4 14B, gpt-oss~20B, and Gemma~4 26B. Llama~3 8B also increases from a matched mean of
0.202 to 0.927 and from zero to three task passes.

The improvement is not universal. Gemma~3 270M increases in mean score from 0.080 to 0.240 but
passes none of the five tasks under either interface. Phi-4 Reasoning Plus is the only deployment
whose matched mean decreases, from 0.309 to 0.080. Three of its five structured responses contain
no parseable JSON object; the remaining two receive partial credit but fail the pass-critical criteria.
The matched comparison therefore shows a large and broadly shared interface benefit, but not a
guaranteed improvement for every model.

\subsection{Structured debugging}
\label{sec:debug}

\begin{figure*}[!t]
\centering
\includegraphics[width=0.96\textwidth]{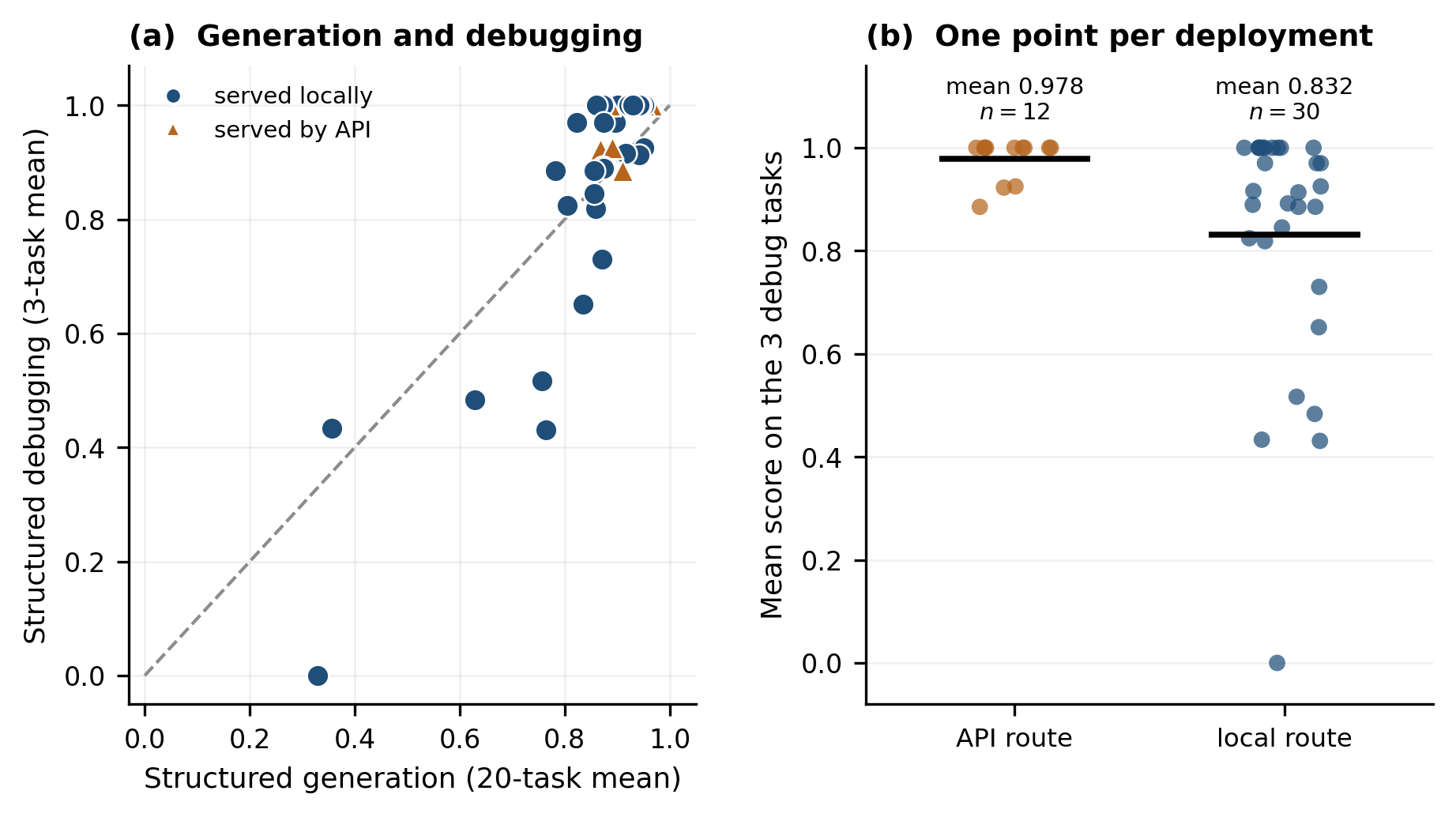}
\caption{Structured generation and debugging. \textbf{(a)} Mean debugging score against the mean
over the twenty schema-described main tasks, with one point per deployment. Circles denote locally
served deployments, triangles denote API-served deployments, and the dashed line marks equal
scores. \textbf{(b)} Debugging means grouped by serving route; horizontal lines show the group
means.}
\label{fig:debug}
\end{figure*}

The three-task extension asks the model to diagnose a supplied faulty artifact and return a JSON
repair patch containing an error location, error class, and corrected content. These tasks do not
test whether a model notices an error in its own earlier response; every deployment diagnoses the
same externally supplied faults.
API-served deployments pass 30 of 36 cases with a mean score of 0.978, while locally served
deployments pass 43 of 90 with a mean score of 0.832. Deployments that perform well on the twenty schema-described main tasks also tend to perform well
on structured debugging. Within this cohort the two scores covary (Spearman $\rho=0.742$,
$p=1.9\times10^{-8}$), although the two evaluated capabilities are not identical
(Fig.~\ref{fig:debug}). A two-sided Mann--Whitney test using one debugging mean per deployment
gives $p=0.010$ for the observed route-group separation. We report these tests as descriptive
summaries of this non-random cohort, not as population-level inference about serving route.
 Deployment-level results are listed in Table~\ref{tab:debug}.

The route comparison is descriptive because model family, size, weight availability, provider, and
serving configuration differ between the two groups. Llama~3.3 70B is the only model version
evaluated through both routes, and it scores 1.000 on all three debugging tasks in both deployments.
This single comparison shows no route difference for that model on these tasks, but it cannot
determine the effect of serving route more generally.
Phi-4 Reasoning Plus is the only deployment that scores zero on all three debugging tasks. In every case, the stored failure labels identify explanatory text but no parseable JSON object,
consistent with its poor structured performance in Sec.~\ref{sec:interface}. Its debugging failure therefore
occurs at the output-contract level: the required repair patch cannot be parsed and evaluated.

\subsection{Run-to-run stability}
\label{sec:stability}

The leaderboard reports one designated run per deployment. To examine how representative a single
run is, we repeated the complete 31-task suite five times for ten locally served deployments whose
designated main-suite scores range from 0.603 to 0.813. Each deployment received the same prompts
under the same serving settings in every repeat. Because the scorer is deterministic, differences
between repeats arise from changes in the returned model artifacts rather than from scoring
randomness.

\begin{figure*}[t]
\centering
\includegraphics[width=0.87\textwidth]{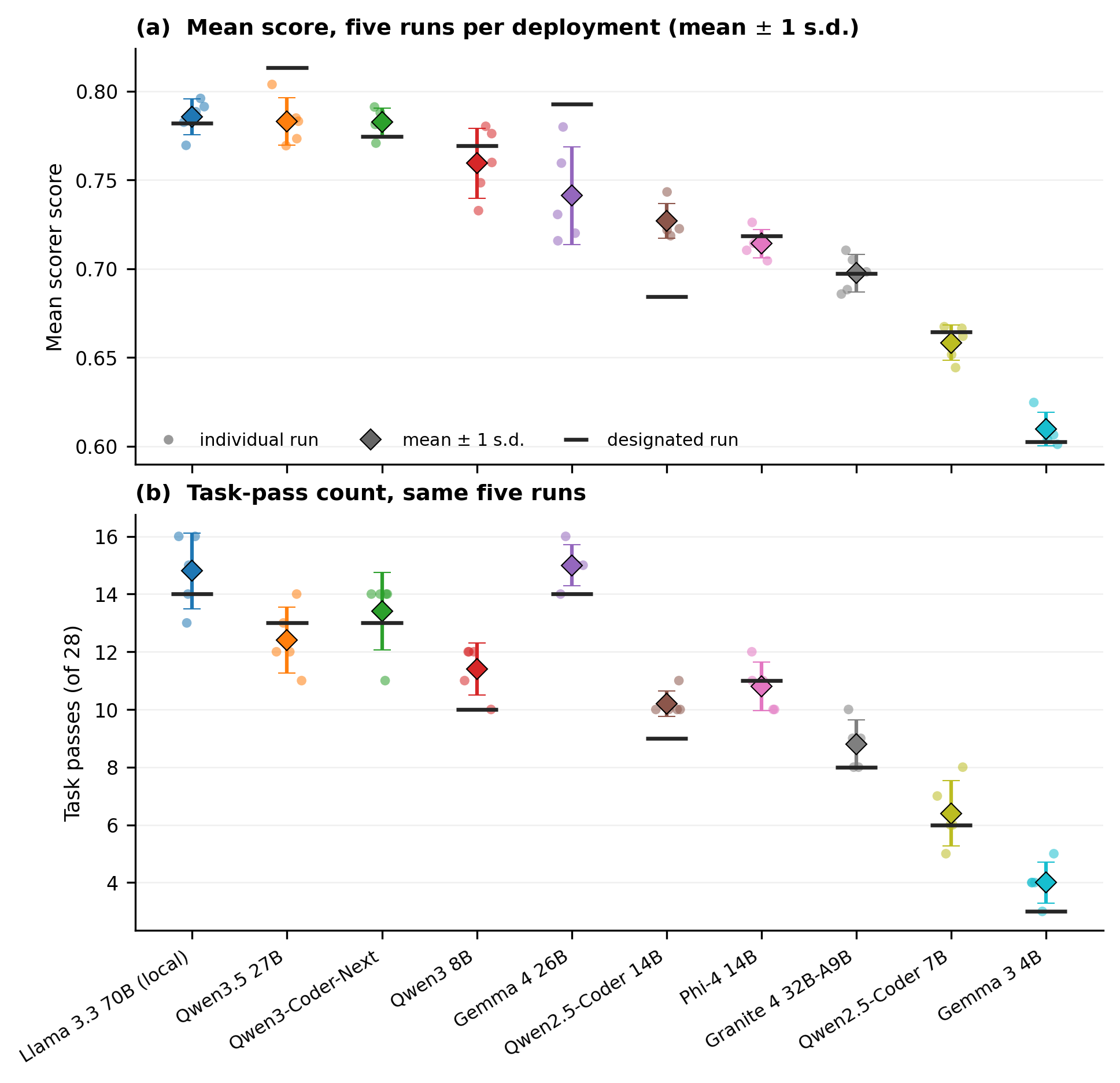}
\caption{Five repeated runs for ten locally served deployments on the 28-task main suite. Light
points show individual runs, diamonds show the mean with one standard deviation, and horizontal
bars show the designated run used in the main results. \textbf{(a)} Mean scorer score.
    \textbf{(b)} Task-pass count. Deployment-level values are given in
Table~\ref{tab:replicates} in Appendix~\ref{app:model-results}.}
\label{fig:stability}
\end{figure*}

The repeat archive was rescored under the same corrected v1.2.1 scorers as the main cohort, so
the values below reflect the B1 and W1 corrections.
On the 28-task main suite, the standard deviation of the five per-run mean scores averages 0.0126
across the ten deployments, indicating little variation in aggregate score. The corresponding
pass-count standard deviation averages 0.94 tasks, and the largest difference between the lowest
and highest pass counts for any deployment is three tasks (Fig.~\ref{fig:stability}). Of the 280
deployment--task combinations, 49 change pass status in at least one repeat, while the remaining
231, or 82.5\%, retain the same pass or fail outcome across all five runs. Thus, most individual
outcomes are stable, although a change to a pass-critical field can flip the pass label without
substantially changing the deployment's mean score.
Gemma~4 26B has the largest run-to-run score standard deviation, 0.0276. For this deployment, 20 of
the 140 main-suite responses reach the configured 4096-token context limit and terminate with
\texttt{length}. These responses are retained as scored model outputs under the benchmark policy
and should be kept in mind when interpreting its wider score variation.

The designated pass count lies within the five-repeat range for nine of the ten deployments.
Qwen2.5-Coder 14B is the exception, with nine designated passes compared with 10--11 across the
repeats. The designated mean score lies within the repeat range for seven deployments; Qwen3.5 27B
and Gemma~4 26B lie above their repeat ranges, while Qwen2.5-Coder 14B lies below. For the ten locally served deployments tested here, the observed run-to-run variation is small
compared with the improvement between the matched native-syntax and structured tasks reported in
Sec.~\ref{sec:interface}. Ordinary variation between repeated runs therefore does not appear large
enough to explain the observed interface effect, and the stability study supports the aggregate
interface result but not fine rankings between deployments separated by only one or two passes. Since the study covers neither
hosted APIs nor every local checkpoint, the measured variability should not be treated as a
universal bound.

\FloatBarrier

\section{Discussion}
\label{sec:discussion-limitations}
\subsection{Implications for HEP software interfaces}
\label{sec:implications}

The matched comparison supports a clear engineering lesson: native HEP tool syntax is a poor place
to leave model uncertainty. A more reliable division of labour is to ask the model for a typed
intermediate representation and let deterministic software validate and render the final artifact.
The schema can enforce required fields, data types, and allowed values, while domain-specific
validators can check particle names, units, and dependencies between settings. The benefit is
particularly large among the 25 deployments below 50 billion parameters, for which task passes
on the matched tasks increase from 2/125 with native syntax to 81/125 with the structured interface,
without task-specific fine-tuning. The complete matched and size-class results are reported in
Table~\ref{tab:pairs}, Fig.~\ref{fig:interface}, and Table~\ref{tab:sizeclass}.
This result does not show that the models ``understand the physics'' and fail only during
transcription. The structured prompts also state the required information more explicitly and
restrict the space of possible responses, and the present experiment does not separate these
effects; isolating the format effect from this added explicitness would require a further ablation,
such as a native-syntax prompt carrying the same explicit list of required fields, which we leave to
future work. The companion system provides indirect evidence on this point: in its matched
three-way comparison, moving four locally served models from direct native generation to a
typed proposal rendered by a builder raises the pass count only from 2/61 to 8/61, whereas adding
deterministic grounding of the user's explicit values and validation raises it to
52/61~\cite{HEPLocalAgent}. Structure alone therefore appears to buy less than the full
schema-mediated pipeline, which is consistent with explicitness and deterministic grounding
carrying much of the benefit measured here. The operational conclusion is narrower: much of the failure observed when models directly
produce native syntax can be avoided when deterministic software owns serialization. Structured
responses can still contain incorrect processes, energies, and settings, so syntactic validity is
not a substitute for domain validation.

The benchmark also distinguishes artifact correctness from broader scientific validity. The
companion system describes five levels at which a workflow can be checked: schema validity,
intent grounding, workflow validity, runtime verification, and scientific
validation~\cite{HEPLocalAgent}. \heptoolbench{} tests the first three: whether the response
parses under the task contract, whether the explicit values of the request survive into the
artifact, and whether the artifact is internally consistent with the released workflow. It does
not execute the artifact and does not judge whether the calculation is meaningful. A response can
therefore be parseable, contract-compliant, and faithful to the stated request without being
executable for the selected perturbative setup or scientifically useful for the intended
analysis. In particular, the
Higgs-plus-jet pair tests artifact construction; the corresponding \proc{p p > h j} request has
no tree-level diagrams in the Standard Model runtime setup used by the companion system and is
not evidence of successful end-to-end Higgs-plus-jet simulation.

The companion system, \textsc{HEPLocalAgent}, applies this design to an executable HEP
workflow~\cite{HEPLocalAgent}. It maps a user's request to a typed representation, validates and
renders the required tool inputs, executes the workflow, and verifies the expected outputs. The
benchmark results provide empirical support for this separation of responsibilities. As a possible
extension, grammar-constrained decoding could enforce the response grammar during
generation~\cite{Willard:2023outlines,Dong:2024xgrammar}; it would complement, rather than replace,
domain validation.
The structured-debugging results in Sec.~\ref{sec:debug} further show why model-generated repairs
must be checked. Although 73 of the 126 repair patches pass all task requirements, the remaining
patches are malformed or contain an incorrect diagnosis or correction. A generated patch should
therefore be treated as a proposed repair rather than applied automatically. The full planner in the
companion agent follows this principle by bounding repair attempts and revalidating the resulting
artifact before execution.

For model selection, parameter count alone is insufficient. Although it is associated with
performance across the evaluated cohort, Fig.~\ref{fig:scale} shows substantial differences between
deployments of similar size. No common external instruction-following or function-calling evaluation
covers the complete cohort, so the present data do not support a quantitative cross-benchmark
predictor. The released HEP-specific tasks provide a more direct test of whether a candidate
deployment satisfies the required artifact contract.

\subsection{Limitations and outlook}
\label{sec:limitations}

The evaluation covers 28 main tasks, while the controlled interface comparison comprises five
matched pairs evaluated across all 42 deployments. Because some tasks share tool conventions and
several deployments belong to related model families, the individual model--task outcomes should
not be treated as independent trials. The statistical results therefore characterize the evaluated
cohort rather than an arbitrary population of models and scientific tasks. Within this scope,
however, the matched improvement reported in Table~\ref{tab:pairs} and
Fig.~\ref{fig:interface} is substantially larger than the run-to-run variation measured in
Sec.~\ref{sec:stability}.
Each task and interface condition uses one fixed prompt. This controls the wording when comparing
the two interfaces, but it does not measure sensitivity to paraphrases, alternative system prompts,
decoding settings, or provider updates. The stability study covers repeated sampling under fixed
settings for ten locally served deployments and should not be interpreted as covering these other
sources of variation. Moreover, a task pass can change when one critical field is incorrect even
if the mean score changes little. Differences of only one or two passes between deployments should
therefore not be regarded as definitive rankings.

The native scorers recognize only equivalences encoded in the benchmark contract. Two of the
five native prompts omitted conventions that their scorers required; those two scorers have been
corrected in v1.2.1 and the whole cohort rescored under them, so this is no longer a caveat on
the reported numbers but a completed correction (Sec.~\ref{sec:task-families}). One asymmetry
does remain in the corrected contract: the structured scorers recover JSON from fenced or
annotated responses while the native scorers score the response verbatim. The audit in
Sec.~\ref{sec:scoring-policy} shows that on the archived subset this asymmetry accounts for a
minority of the native failures and that the structured advantage survives under symmetric
extraction, but it also shows that roughly half of the structured passes rely on tolerant
extraction. The headline 21/210 and 159/210 counts should therefore be read as the corrected
contract's view under an extraction rule that still differs between the two conditions, rather
than as extraction-neutral quantities. A future release should apply the same extraction rule on
both sides and report both the verbatim and the
tolerant view; requiring bare JSON or using grammar-constrained decoding would define a further,
stricter interface condition. Because the audit covers only locally served deployments, and the
hosted deployments account for most of the native passes, a full-cohort rescoring under a single
symmetric extraction rule is the most important missing measurement.
The tested HEP tools have public documentation that may have appeared in model pretraining data.
Prior familiarity could improve absolute performance relative to an unfamiliar scientific
package~\cite{Sainz:2023contamination,Xu:2024contamination}, although it does not imply that the
benchmark tasks themselves were present in the training data. It also does not explain the matched
interface improvement, since both versions of each request use the same deployment and describe the
same physical workflow. However, the absolute performance measured for these well-documented HEP tools may not generalize
to new scientific software that was absent from the models' training data.

Finally, parameter count, model family, quantization, provider, and serving route are not
independently controlled. The association with model scale in Fig.~\ref{fig:scale} is therefore
observational, and the single Llama~3.3 70B route comparison cannot establish a general route
effect or its absence. Hosted aliases may change over time, while locally served outputs depend on
the selected weights, quantization, and runtime. The released results should consequently be read
as a dated, configuration-specific snapshot rather than permanent model attributes.

Natural extensions include additional matched requests, prompt variants, repeated API evaluations,
controlled changes to schema detail and decoding constraints, and broader HEP workflows. NLO setup,
richer decay chains, parameter-card scans, recasting frameworks, spectrum generators, and EFT tools
can all be evaluated using the same basic unit: a scientific request, a concrete artifact, and a
deterministic definition of whether that artifact is usable.

\section{Conclusion}
\label{sec:conclusion}
\heptoolbench{} evaluates whether a single model response satisfies a deterministic artifact
contract for HEP software. Across five matched requests, the schema-mediated interface raises the
mean score from 0.418 to 0.902 and the task-pass count from 21/210 to 159/210, with 41 of 42
deployments improving. Many failures in direct native-syntax generation can therefore be avoided
when the model supplies typed fields and deterministic software owns serialization.

The effect size is specific to the corrected v1.2.1 benchmark contract. The structured condition
combines field decomposition, tolerant extraction, and deterministic rendering. The two native
scorers that had enforced operational conventions their own prompts never stated have been
rescored prompt-faithfully across the cohort, so the native side reported here is the
prompt-faithful one; the native scorers still score responses verbatim, and that extraction
asymmetry remains. An audit of archived responses shows that it changes the size
of the effect but not its presence: under a common extraction rule the structured interface still
passes several times more often. A task pass establishes compliance under the implemented
checks, not runtime or scientific validity. The result is therefore an engineering comparison of two
interfaces, not proof that every passing calculation will execute successfully. Typed intermediate
representations nevertheless provide a more reliable foundation for guarded workflows, including
the companion \textsc{HEPLocalAgent} system~\cite{HEPLocalAgent}. The tasks, scorers, responses,
and benchmark code are released for independent checking and extension~\cite{HEPToolBenchRepo};
the result CSVs used for the reported summaries accompany the manuscript source.

\section*{Acknowledgements}

A.S. acknowledges the financial support and institutional resources provided by the Indian
Institute of Science (IISc). S.K.V. acknowledges the receipt of REDA funds from IISc.

\section*{Code and data availability}
\label{sec:availability}
The benchmark definitions, prompts, deterministic scorers, runners, and curated result artifacts
are available at \url{https://github.com/AadarshSingh0/HEPToolBench}~\cite{HEPToolBenchRepo}.
Release v1.2, which produced the original cohort run, is permanently archived on
Zenodo~\cite{singh2026heptoolbench} and is retained there unchanged for historical
reproducibility.

The results reported in this paper are those of the corrected candidate v1.2.1, scoring policy
\texttt{B1\_W1\_S1R}. Relative to v1.2 it changes two native scorers, so that
\texttt{mg\_basic\_002} accepts any syntactically valid output-directory name and
\texttt{mg\_workflow\_005} does not charge the unstated \texttt{analysis=OFF} and \texttt{done}
conventions, and it replaces four Sarvam-105B provider-rejected calls with repeated measurements
under the same request configuration (Sec.~\ref{sec:scoring-policy}). All prompts, task
definitions, and every other scorer are unchanged. At the time of writing, v1.2.1 has not yet
been tagged, released, or deposited; the corrected scorers, the regeneration scripts, and the
corrected result files will be published under a v1.2.1 tag with its own Zenodo deposit, and the
archived v1.2 record will remain available alongside it.

The manuscript source contains the corrected v1.2.1 result CSV for all 42 deployments
(1,302 records: 42 deployments $\times$ 31 tasks) and the separate
1550-row five-repeat stability CSV used for the reported summaries. The convention-and-extraction
audit in Sec.~\ref{sec:scoring-policy} is a secondary analysis of the archived locally served
responses and the stability archive, carried out against the v1.2 scorers; its aggregate counts
and scope are reported in the paper, and per-unit audit artifacts are not included in the
manuscript source.
The consolidated CSV is the analysis snapshot used in this paper. It retains the source run
identifiers, infrastructure status, runtime configuration, per-record correction provenance, and
paths to the underlying response records. Model aliases and configurations are reported as
evaluated; hosted aliases may change over time.

\section*{Declaration of AI-assisted development and writing}

During the preparation of this work, the authors used OpenAI ChatGPT and Anthropic
Claude to assist with manuscript organization, drafting, and language refinement.
The authors reviewed and edited all AI-assisted content and take full responsibility
for the content of the article.

\appendix

\section{Task definitions and scoring details}
\label{app:tasks-scoring}
\begingroup
\footnotesize

\begin{longtable}{
@{}>{\ttfamily\raggedright\arraybackslash}p{0.30\textwidth}
>{\raggedright\arraybackslash}p{0.34\textwidth}
>{\raggedright\arraybackslash}p{0.25\textwidth}@{}
}
\caption{Task composition of \heptoolbench{} v1.2, unchanged in the corrected v1.2.1. The first
28 tasks constitute the main benchmark, and the final three constitute the structured-debugging
extension.}
\label{tab:task-list-full}\\

\toprule
Task ID & Family & Output type \\
\midrule
\endfirsthead

\multicolumn{3}{@{}l}{\tablename~\thetable\ continued}\\
\toprule
Task ID & Family & Output type \\
\midrule
\endhead

\midrule
\multicolumn{3}{r@{}}{\emph{Continued on next page}}\\
\endfoot

\bottomrule
\endlastfoot

mg\_basic\_001 & MadGraph process generation & native process card \\
mg\_debug\_001 & MadGraph debugging & repaired process card \\
mg\_structured\_001 & MadGraph process generation & schema-described JSON \\

mg\_basic\_002 & MadGraph process generation & native process card \\
mg\_debug\_002 & MadGraph debugging & repaired process card \\
mg\_structured\_002 & MadGraph process generation & schema-described JSON \\

mg\_basic\_003 & MadGraph process generation & native process card \\
mg\_debug\_003 & MadGraph debugging & repaired process card \\
mg\_structured\_003 & MadGraph process generation & schema-described JSON \\

mg\_runcard\_004 & Run-card configuration & native run card \\
mg\_runcard\_structured\_004 & Run-card configuration & schema-described JSON \\

mg\_workflow\_005 & MG5/Pythia8/Delphes workflow & native workflow script \\
mg\_workflow\_structured\_005 & MG5/Pythia8/Delphes workflow & schema-described JSON \\

mg\_parse\_006 & MadGraph log parsing & schema-described JSON \\
mg\_parse\_007 & MadGraph failure diagnosis & schema-described JSON \\
mg\_parse\_008 & Cross-section unit conversion & schema-described JSON \\
mg\_parse\_009 & Output validation & schema-described JSON \\
pythia\_config\_010 & Pythia8 configuration validation & schema-described JSON \\
delphes\_objects\_011 & Delphes object validation & schema-described JSON \\
lhe\_sanity\_012 & LHE sanity checks & schema-described JSON \\
cutflow\_diagnosis\_013 & Cut-flow diagnosis & schema-described JSON \\
scan\_plan\_014 & Parameter-scan planning & schema-described JSON \\
param\_card\_patch\_015 & Parameter-card patching & schema-described JSON \\
scan\_results\_016 & Scan-result summarization & schema-described JSON \\
scan\_recovery\_017 & Scan-recovery planning & schema-described JSON \\
benchmark\_recommendation\_018 & Benchmark-point recommendation & schema-described JSON \\
plot\_data\_019 & Plot-data preparation & schema-described JSON \\
repro\_audit\_020 & Reproducibility audit & schema-described JSON \\

mg\_debug\_structured\_001 & Structured debugging extension & JSON repair patch \\
mg\_debug\_structured\_002 & Structured debugging extension & JSON repair patch \\
mg\_debug\_structured\_003 & Structured debugging extension & JSON repair patch \\

\end{longtable}
\endgroup
Each task records a continuous score in $[0,1]$ and a binary pass label. The score awards
partial credit to task-specific components, whereas a pass requires the conditions defined as
critical under the task's benchmark contract. There is no universal score
threshold for passing: in the curated main-suite snapshot, the highest failing score is 0.96 and
the lowest passing score is 0.79. On 14 of the 28 main tasks, however, the lowest observed passing
score is at least 0.95.

The task-pass counts reported throughout this paper use the scorer field \texttt{passed}. The
structured scorers, including all five used in the matched pairs, also record an auxiliary field
named \texttt{strict\_passed}, which additionally requires bare JSON without a Markdown fence or
surrounding explanation. We reserve the term \emph{strict-format pass} for that auxiliary field;
it is not used for the task-pass counts reported
in the paper; Sec.~\ref{sec:scoring-policy} reports how many matched-pair passes it would remove.

For native-syntax tasks, the scorer checks the returned card, command sequence, or repair directly.
Only equivalences explicitly encoded in the scorer are normalized. Missing commands, invalid
particle tokens, incorrect beam energies, output paths, or workflow settings reduce the partial
score and may violate pass-critical requirements. An invalid card is not rewritten before it is
scored.
The native contract also contains fixed launch and bookkeeping conventions, and semantically
equivalent alternatives not explicitly encoded by the scorer can therefore fail. Two conventions
that the corresponding prompts never stated were removed in v1.2.1: \texttt{mg\_basic\_002} now
accepts any syntactically valid \texttt{output <directory>} command rather than one fixed
reference name, and \texttt{mg\_workflow\_005} no longer treats \texttt{analysis=OFF} or a
terminating \texttt{done} as pass-critical or score-bearing. Both remain visible in the scorer
diagnostics. Every other requirement in both tasks, and every other native scorer, is unchanged;
\texttt{mg\_debug\_002}, which delegates to the basic scorer, explicitly retains the historical
reference-name behaviour so that the debugging contract is not altered.
In addition, the native scorers check the recorded response text as returned: a Markdown code
fence or a visible reasoning block anywhere in the response is a hard failure, and the native
prompts for the process and run-card tasks ask only for no explanation rather than explicitly
forbidding fences. The native pass rate therefore measures direct consumability of the verbatim
response by the released workflow, not equivalence under every valid HEP-tool representation.

Each structured task uses a fixed extraction routine followed by deterministic semantic checks.
These routines may recover a JSON object from a Markdown fence or surrounding text while recording
the corresponding format deviation. A small number of analysis-oriented tasks can also recover
explicitly named fields from otherwise legible JSON-like output. The rules for each task are
applied unchanged to every deployment; no model-specific parsing exceptions are used. Where
applicable, the recovered fields are passed to a deterministic builder, and the resulting card or
workflow is also validated.

The structured prompts provide the artifact shape without pre-filling the serialized answer. For
example, a value-filled template would directly provide entries such as

\begin{lstlisting}
{
  "model": "sm",
  "initial_state": ["p", "p"],
  "final_state": ["e+", "e-"],
  "beam_energy_gev": 6500
}
\end{lstlisting}

whereas the benchmark supplies placeholders:

\begin{lstlisting}
{
  "model": "<string>",
  "initial_state": ["<particle>", "<particle>"],
  "final_state": ["<particle>", "<particle>"],
  "beam_energy_gev": "<number>"
}
\end{lstlisting}

The physical process and collider energy are stated in the prose request, as they are for the
matched native-syntax task. The distinction is therefore not that the target values are secret,
but that the schema specifies the required structure without filling those values into the
artifact for the model.

Infrastructure validity is assessed separately from artifact quality. Provider quota failures,
server errors, timeouts, missing responses, and runner exceptions are not treated as completed
model answers and are excluded or repeated under the clean-run policy. By contrast, a malformed
artifact returned by a completed inference call remains a valid measurement and is scored as a
model failure. A deployment enters the final cohort only after a valid record is available for
every expected task. The four Sarvam-105B calls rejected by the provider before generation were
handled under this rule and repeated under an unchanged request configuration
(Sec.~\ref{sec:scoring-policy}); no HTTP-error placeholder remains in the released records.

The consolidated main-result CSV is the curated v1.2.1 snapshot used by the analysis and
regeneration scripts. It contains the final validity flags, selected source identifiers, runtime
metadata, per-record correction provenance, and recorded paths to response and result artifacts. The benchmark runner creates richer
per-run JSON records for new executions. The paper figures and tables should therefore be
regenerated from the released consolidated snapshot rather than by interpreting every raw zero in
an unfiltered run directory as a completed model response.

\clearpage
\section{Supplementary model results}
\label{app:model-results}
\begingroup
\footnotesize
\setlength{\tabcolsep}{3.2pt}
\renewcommand{\arraystretch}{0.93}

\begin{longtable}{@{}llllrrcrrr@{}}

\caption{Results for all 42 deployments with valid records for the 28 main-suite tasks and
the three structured-debugging tasks, ordered by mean main-suite score, under the corrected
v1.2.1 \texttt{B1\_W1\_S1R} contract. Weights indicates
whether checkpoint weights are publicly downloadable. Params gives the total parameter
count in billions; active parameters are shown in parentheses for mixture-of-experts models,
and \emph{n/d} denotes an undisclosed size. Pass and Mean are the number of fully passed tasks and the mean partial-credit
score on the main suite; the Wilson 95\% confidence interval is for the pass
rate, Pass/28. Native and Struct.\ report passes on its eight native-syntax
and 20 schema-described tasks, respectively; Dbg.\ reports passes on the three
structured-debugging tasks.}
\label{tab:leaderboard}\\

\toprule
Model & Route & Weights & Params & Pass & Mean & 95\% CI & Native & Struct. & Dbg. \\
\midrule
\endfirsthead

\multicolumn{10}{@{}l}{\tablename~\thetable\ (continued)}\\
\toprule
Model & Route & Weights & Params & Pass & Mean & 95\% CI & Native & Struct. & Dbg. \\
\midrule
\endhead

\midrule
\multicolumn{10}{r@{}}{\emph{Continued on next page}}\\
\endfoot

\bottomrule
\endlastfoot

\texttt{GPT-4.1} & API & closed & n/d & 19/28 & 0.945 & [0.49, 0.82] & 5/8 & 14/20 & 3/3 \\
\texttt{gpt-oss 120B} & local & open & 120 (5.1) & 19/28 & 0.936 & [0.49, 0.82] & 5/8 & 14/20 & 3/3 \\
\texttt{Gemini 3.1 Flash-Lite} & API & closed & n/d & 20/28 & 0.934 & [0.53, 0.85] & 6/8 & 14/20 & 3/3 \\
\texttt{GPT-4.1 mini} & API & closed & n/d & 18/28 & 0.928 & [0.46, 0.79] & 4/8 & 14/20 & 3/3 \\
\texttt{Qwen3-Next 80B-A3B} & local & open & 80 (3.0) & 16/28 & 0.858 & [0.39, 0.73] & 3/8 & 13/20 & 2/3 \\
\texttt{Gemini 3.5 Flash} & API & closed & n/d & 16/28 & 0.854 & [0.39, 0.73] & 1/8 & 15/20 & 3/3 \\
\texttt{gpt-oss 20B} & local & open & 20 (3.6) & 14/28 & 0.840 & [0.33, 0.67] & 0/8 & 14/20 & 3/3 \\
\texttt{Qwen3 14B} & local & open & 14 & 14/28 & 0.815 & [0.33, 0.67] & 0/8 & 14/20 & 3/3 \\
\texttt{Qwen3.5 27B} & local & open & 27 & 13/28 & 0.813 & [0.30, 0.64] & 2/8 & 11/20 & 1/3 \\
\texttt{Llama 3.3 70B (API)} & API & open & 70 & 16/28 & 0.812 & [0.39, 0.73] & 1/8 & 15/20 & 3/3 \\
\texttt{GPT-4.1 nano} & API & closed & n/d & 13/28 & 0.806 & [0.30, 0.64] & 1/8 & 12/20 & 0/3 \\
\texttt{Gemma 4 26B} & local & open & 26 & 14/28 & 0.793 & [0.33, 0.67] & 0/8 & 14/20 & 3/3 \\
\texttt{Magistral Medium 2509} & API & closed & n/d & 15/28 & 0.788 & [0.36, 0.70] & 0/8 & 15/20 & 3/3 \\
\texttt{Devstral 2512} & API & open & 123 & 12/28 & 0.786 & [0.27, 0.61] & 0/8 & 12/20 & 3/3 \\
\texttt{Sarvam 105B} & API & open & 105 (10.3) & 12/28 & 0.784 & [0.27, 0.61] & 0/8 & 12/20 & 2/3 \\
\texttt{Mistral Medium 2508} & API & closed & n/d & 13/28 & 0.783 & [0.30, 0.64] & 0/8 & 13/20 & 3/3 \\
\texttt{Llama 3.3 70B (local)} & local & open & 70 & 14/28 & 0.782 & [0.33, 0.67] & 0/8 & 14/20 & 3/3 \\
\texttt{Gemma 4 31B} & local & open & 31 & 14/28 & 0.779 & [0.33, 0.67] & 0/8 & 14/20 & 2/3 \\
\texttt{Qwen3-Coder-Next} & local & open & 80 (3.0) & 13/28 & 0.774 & [0.30, 0.64] & 1/8 & 12/20 & 2/3 \\
\texttt{Mistral Large 2512} & API & open & 675 (41.0) & 13/28 & 0.771 & [0.30, 0.64] & 0/8 & 13/20 & 3/3 \\
\texttt{Qwen3 8B} & local & open & 8 & 10/28 & 0.769 & [0.21, 0.54] & 0/8 & 10/20 & 3/3 \\
\texttt{Codestral 2508} & API & closed & n/d & 11/28 & 0.767 & [0.24, 0.58] & 0/8 & 11/20 & 1/3 \\
\texttt{Qwen3.5 35B} & local & open & 35 & 8/28 & 0.755 & [0.15, 0.47] & 1/8 & 7/20 & 2/3 \\
\texttt{Gemma 4 12B} & local & open & 12 & 12/28 & 0.753 & [0.27, 0.61] & 0/8 & 12/20 & 3/3 \\
\texttt{DeepSeek-R1 70B} & local & open & 70 & 13/28 & 0.750 & [0.30, 0.64] & 0/8 & 13/20 & 3/3 \\
\texttt{Devstral 24B} & local & open & 24 & 10/28 & 0.739 & [0.21, 0.54] & 0/8 & 10/20 & 3/3 \\
\texttt{Mistral Small 3.2 24B} & local & open & 24 & 8/28 & 0.733 & [0.15, 0.47] & 0/8 & 8/20 & 3/3 \\
\texttt{Phi-4 14B} & local & open & 14 & 11/28 & 0.718 & [0.24, 0.58] & 0/8 & 11/20 & 1/3 \\
\texttt{Gemma 3 12B} & local & open & 12 & 11/28 & 0.712 & [0.24, 0.58] & 0/8 & 11/20 & 1/3 \\
\texttt{Ministral 3 14B} & local & open & 14 & 9/28 & 0.706 & [0.18, 0.51] & 0/8 & 9/20 & 0/3 \\
\texttt{Gemma 4 E4B} & local & open & 4 & 9/28 & 0.699 & [0.18, 0.51] & 0/8 & 9/20 & 0/3 \\
\texttt{Granite 4 32B-A9B} & local & open & 32 (9.0) & 8/28 & 0.697 & [0.15, 0.47] & 0/8 & 8/20 & 0/3 \\
\texttt{Gemma 4 E2B} & local & open & 2 & 6/28 & 0.689 & [0.10, 0.40] & 0/8 & 6/20 & 0/3 \\
\texttt{Falcon 3 10B} & local & open & 10 & 11/28 & 0.685 & [0.24, 0.58] & 0/8 & 11/20 & 0/3 \\
\texttt{Qwen2.5-Coder 14B} & local & open & 14 & 9/28 & 0.684 & [0.18, 0.51] & 0/8 & 9/20 & 2/3 \\
\texttt{Qwen2.5-Coder 7B} & local & open & 7 & 6/28 & 0.664 & [0.10, 0.40] & 0/8 & 6/20 & 0/3 \\
\texttt{Llama 3 8B} & local & open & 8 & 6/28 & 0.657 & [0.10, 0.40] & 0/8 & 6/20 & 0/3 \\
\texttt{Llama 3.1 8B} & local & open & 8 & 6/28 & 0.625 & [0.10, 0.40] & 0/8 & 6/20 & 0/3 \\
\texttt{Gemma 3 4B} & local & open & 4 & 3/28 & 0.603 & [0.04, 0.27] & 0/8 & 3/20 & 0/3 \\
\texttt{Phi-4 mini 3.8B} & local & open & 4 & 3/28 & 0.488 & [0.04, 0.27] & 0/8 & 3/20 & 0/3 \\
\texttt{Phi-4 Reasoning Plus} & local & open & 14 & 2/28 & 0.336 & [0.02, 0.23] & 0/8 & 2/20 & 0/3 \\
\texttt{Gemma 3 270M} & local & open & 0.27 & 1/28 & 0.277 & [0.01, 0.18] & 0/8 & 1/20 & 0/3 \\

\end{longtable}
\endgroup

\begin{table}[p]
\centering
\footnotesize
\renewcommand{\arraystretch}{0.90}

\caption{Mean scores and passes on the twenty schema-described main-suite tasks and the
three-task structured-debugging extension, grouped by serving route, under the corrected
v1.2.1 \texttt{B1\_W1\_S1R} contract. The grouping is
descriptive: route is confounded with model identity, public weight availability, and model size.}
\label{tab:debug}

\begin{tabular}{@{}lrrrr@{}}
\toprule
& \multicolumn{2}{c}{Schema-described main tasks}
& \multicolumn{2}{c}{Debugging extension} \\
\cmidrule(lr){2-3}
\cmidrule(lr){4-5}
Model & Mean & Passes & Mean & Passes \\
\midrule
\multicolumn{5}{l}{\emph{API-served}}\\
\texttt{Gemini 3.1 Flash-Lite} & 0.952 & 14/20 & 1.000 & 3/3 \\
\texttt{GPT-4.1} & 0.966 & 14/20 & 1.000 & 3/3 \\
\texttt{GPT-4.1 mini} & 0.965 & 14/20 & 1.000 & 3/3 \\
\texttt{Gemini 3.5 Flash} & 0.951 & 15/20 & 1.000 & 3/3 \\
\texttt{Llama 3.3 70B (API)} & 0.895 & 15/20 & 1.000 & 3/3 \\
\texttt{Magistral Medium 2509} & 0.928 & 15/20 & 1.000 & 3/3 \\
\texttt{Mistral Medium 2508} & 0.885 & 13/20 & 1.000 & 3/3 \\
\texttt{Mistral Large 2512} & 0.879 & 13/20 & 1.000 & 3/3 \\
\texttt{Devstral 2512} & 0.886 & 12/20 & 1.000 & 3/3 \\
\texttt{Sarvam 105B} & 0.938 & 12/20 & 0.925 & 2/3 \\
\texttt{Codestral 2508} & 0.869 & 11/20 & 0.923 & 1/3 \\
\texttt{GPT-4.1 nano} & 0.910 & 12/20 & 0.885 & 0/3 \\
\midrule
\multicolumn{5}{l}{\emph{Served locally through Ollama}}\\
\texttt{gpt-oss 120B} & 0.957 & 14/20 & 1.000 & 3/3 \\
\texttt{gpt-oss 20B} & 0.951 & 14/20 & 1.000 & 3/3 \\
\texttt{Gemma 4 26B} & 0.942 & 14/20 & 1.000 & 3/3 \\
\texttt{Qwen3 14B} & 0.946 & 14/20 & 1.000 & 3/3 \\
\texttt{Llama 3.3 70B (local)} & 0.900 & 14/20 & 1.000 & 3/3 \\
\texttt{DeepSeek-R1 70B} & 0.926 & 13/20 & 1.000 & 3/3 \\
\texttt{Gemma 4 12B} & 0.922 & 12/20 & 1.000 & 3/3 \\
\texttt{Mistral Small 3.2 24B} & 0.861 & 8/20 & 1.000 & 3/3 \\
\texttt{Devstral 24B} & 0.872 & 10/20 & 1.000 & 3/3 \\
\texttt{Qwen3 8B} & 0.929 & 10/20 & 1.000 & 3/3 \\
\texttt{Qwen2.5-Coder 14B} & 0.823 & 9/20 & 0.970 & 2/3 \\
\texttt{Qwen3-Coder-Next} & 0.897 & 12/20 & 0.970 & 2/3 \\
\texttt{Qwen3.5 35B} & 0.874 & 7/20 & 0.970 & 2/3 \\
\texttt{Qwen3-Next 80B-A3B} & 0.951 & 13/20 & 0.925 & 2/3 \\
\texttt{Qwen3.5 27B} & 0.915 & 11/20 & 0.916 & 1/3 \\
\texttt{Gemma 4 31B} & 0.941 & 14/20 & 0.913 & 2/3 \\
\texttt{Phi-4 14B} & 0.872 & 11/20 & 0.892 & 1/3 \\
\texttt{Gemma 3 12B} & 0.874 & 11/20 & 0.889 & 1/3 \\
\texttt{Qwen2.5-Coder 7B} & 0.783 & 6/20 & 0.885 & 0/3 \\
\texttt{Granite 4 32B-A9B} & 0.855 & 8/20 & 0.885 & 0/3 \\
\texttt{Ministral 3 14B} & 0.856 & 9/20 & 0.845 & 0/3 \\
\texttt{Llama 3 8B} & 0.804 & 6/20 & 0.824 & 0/3 \\
\texttt{Falcon 3 10B} & 0.859 & 11/20 & 0.819 & 0/3 \\
\texttt{Gemma 4 E4B} & 0.871 & 9/20 & 0.730 & 0/3 \\
\texttt{Gemma 4 E2B} & 0.835 & 6/20 & 0.652 & 0/3 \\
\texttt{Gemma 3 4B} & 0.757 & 3/20 & 0.517 & 0/3 \\
\texttt{Phi-4 mini 3.8B} & 0.629 & 3/20 & 0.483 & 0/3 \\
\texttt{Gemma 3 270M} & 0.356 & 1/20 & 0.433 & 0/3 \\
\texttt{Llama 3.1 8B} & 0.763 & 6/20 & 0.431 & 0/3 \\
\texttt{Phi-4 Reasoning Plus} & 0.330 & 2/20 & 0.000 & 0/3 \\
\bottomrule
\end{tabular}
\end{table}

\begin{table}[t]
\centering
\small
\caption{Matched-interface and structured-debugging results by size class, under the
corrected v1.2.1 \texttt{B1\_W1\_S1R} contract.
$N$ is the number of deployments. Native and Struct.\ aggregate the five matched
task pairs, while Dbg.\ aggregates the three structured-debugging tasks.
The 50\,B threshold uses total parameter count and is a presentational convention
(Sec.~\ref{sec:cohort}); Fig.~\ref{fig:scale} shows the corresponding continuous
relationship. Differences are calculated from the unrounded mean scores.}
\label{tab:sizeclass}

\begin{tabular}{@{}lrrrrccc@{}}
\toprule
& & \multicolumn{3}{c}{Paired mean score}
& \multicolumn{2}{c}{Paired passes} & Dbg. \\
\cmidrule(lr){3-5}
\cmidrule(lr){6-7}
Class & $N$ & Native & Struct. & $\Delta$ & Native & Struct. & passes \\
\midrule
Disclosed $<$50\,B
& 25 & 0.292 & 0.853 & $+0.561$ & 2/125 & 81/125 & 30/75 \\

Disclosed $\geq$50\,B
& 9 & 0.536 & 0.967 & $+0.431$ & 7/45 & 40/45 & 24/27 \\

Size undisclosed
& 8 & 0.680 & 0.980 & $+0.300$ & 12/40 & 38/40 & 19/24 \\
\bottomrule
\end{tabular}
\end{table}

\begin{table}[t]
\centering
\small
\caption{Run-to-run stability on the 28-task main suite for ten locally served
deployments, ordered by mean pass count, under the corrected v1.2.1
\texttt{B1\_W1\_S1R} scorers. Each deployment was evaluated five times
with its serving configuration held fixed. Pass-count and mean-score SDs are sample
standard deviations across the five repeats; the final row reports their mean across
deployments.}
\label{tab:replicates}

\begin{tabular}{@{}lccccc@{}}
\toprule
Deployment & Mean pass count & Pass range & Pass SD & Mean score & Score SD \\
\midrule
Gemma 4 26B & 15.0 & 14--16 & 0.71 & 0.741 & 0.0276 \\
Llama 3.3 70B (local) & 14.8 & 13--16 & 1.30 & 0.785 & 0.0102 \\
Qwen3-Coder-Next & 13.4 & 11--14 & 1.34 & 0.783 & 0.0078 \\
Qwen3.5 27B & 12.4 & 11--14 & 1.14 & 0.783 & 0.0134 \\
Qwen3 8B & 11.4 & 10--12 & 0.89 & 0.759 & 0.0197 \\
Phi-4 14B & 10.8 & 10--12 & 0.84 & 0.714 & 0.0079 \\
Qwen2.5-Coder 14B & 10.2 & 10--11 & 0.45 & 0.727 & 0.0098 \\
Granite 4 32B-A9B & 8.8 & 8--10 & 0.84 & 0.698 & 0.0106 \\
Qwen2.5-Coder 7B & 6.4 & 5--8 & 1.14 & 0.658 & 0.0100 \\
Gemma 3 4B & 4.0 & 3--5 & 0.71 & 0.610 & 0.0093 \\
\midrule
\multicolumn{3}{@{}l}{Mean deployment-level SD} & 0.94 & & 0.0126 \\
\bottomrule
\end{tabular}
\end{table}
\clearpage

\bibliographystyle{unsrtnat}
\bibliography{refs}

\end{document}